\documentclass[journal=jctcce,manuscript=article]{achemso}
\setkeys{acs}{keywords=true}
\usepackage[utf8]{inputenc}
\usepackage{caption}
\usepackage{amsmath,subdepth,bm}
\usepackage{csquotes}
\usepackage{xcolor}
\usepackage{siunitx}
\usepackage{upgreek}
\usepackage{chemmacros}
\setcitestyle{round,numbers}

\chemsetup{
  modules=units,
}
\DeclareSIUnit{\kcalmol}{\kilo\cal\per\mole}
\DeclareSIUnit{\angstrom}{\text{Å}}
\DeclareSIUnit{\hartree}{\text{\ensuremath{E}}_{\mathrm{h}}}

\MakeOuterQuote{"}

\title{Multiconfigurational quantum chemistry:\\The CASPT2 method}
\author{Stefano Battaglia, Ignacio Fdez. Galván, and Roland Lindh}
\affiliation{
  Department of Chemistry - BMC, Uppsala University,
  P.O. Box 576, SE-75123 Uppsala, Sweden
}
\email{stefano.battaglia@kemi.uu.se, ignacio.fernandez@kemi.uu.se, roland.lindh@kemi.uu.se}
\date{February 2022}


\begin{document}

\maketitle
\begin{abstract}
This chapter presents the theory behind the CASPT2 method and its adaptation to a multi-state formalism. The chapter starts with an introduction of the theory of the CASPT2 method -- an application of Rayleigh--Schrödinger perturbation theory applied to multiconfigurational reference function -- as it was originally presented. In particular, we discuss the nature of the reference Hamiltonian and the first-order interacting space. This is followed by some detailed discussion with respect to the intruder state problem and various shift techniques to address this problem. Afterwards a longer review on alternative reference Hamiltonians, which to some degree or completely remove the intruder state problem, is put forward. Subsequently the presently proposed multi-state versions of the CASPT2 method are discussed in some detail. The chapter is concluded with a review of different benchmark assessments of the accuracy of the method and a qualified suggestion on the future development potentials of the approach.  
\end{abstract}

\section{Introduction}
The community of theoretical chemists had made during the 1960s tremendous progress on simulating properties of ground state molecules made up of the elements of the upper part of the periodic table. Still, however, a lot of
problems remained to be solved using techniques with explicit wave functions models.
In particular, single-reference-state methods -- where the primary building block is a single electron configuration as represented by a Slater determinant -- had problems with, for example, bond breaking and the properties
of transition metal, actinides, and lanthanides. It turned out that the problem was two-fold.
The first was the intrinsic problem of the near-degeneracy that manifests itself in, for example, bond-breaking processes where bonding and anti-bonding orbitals go towards energetic degeneracy, or in the case of d- and f-elements with partial occupation in which a multitude of
different electronic configurations are degenerate or very nearly degenerate. A solution to this particular near-degeneracy problem
was the development in the 1970s of the so-called multi-configurational approaches as expressed by the now famous CASSCF \cite{roos1980complete,siegbahn1980comparison,siegbahn1981complete} and the FORS methods \cite{ruedenberg1982atoms} (for an excellent review on the methods read the 2011 paper by J. Olsen \cite{olsen2011casscf} or in the book "Multiconfigurational Quantum Chemistry" \cite{Roos2016}).
The second problem was the issue of a proper description of the electron--electron correlation. There where methods like
the multi-reference CI method (MRCI) of P. E. M. Siegbahn \cite{siegbahn1979generalizations,siegbahn1980generalizations}, a multitude of single-reference methods such as single and double CI (CISD) method (consult the book by Shavitt for an assessment of the state of the art in 1977 \cite{shavitt1977modern}), Møller--Plesset second order perturbation theory (MP2) applied to a Hartree--Fock single-determinant reference function \cite{moller1934note}, and the coupled-cluster singles and doubles (CCSD) \cite{purvis1982full} that addressed these issues. For a report on the status at the time of perturbation theory and coupled-cluster methods read the review by Bartlett \cite{bartlett1981many}. While the MRCI and CISD approaches were problematic due to improper scaling with system size -- lack of size consistency and size extensivity -- there were also qualitative issues with MP2 and CCSD due to their single-reference nature -- both methods, for example, tend to diverge or produce erroneous energetics at bond breaking.
The research group of the late Prof. Björn O. Roos, at Lund University, looked for a solution to the problem and found it natural that
a possible road forward was to apply second order perturbation theory to a contracted CASSCF reference function. 
Initial investigations in 1982 did not, however, give comforting results \cite{roos1982simple}. 
The investment of resources into the project -- designating a Ph.D. student, Kerstin Andersson, to particularly address the issue -- finally resulted in progress.
In 1990, the first paper was presented which introduced the CASPT2 method \cite{andersson1990second}, followed by a second one in 1992 \cite{andersson1992second}.
The key difference to the 1982 investigation was to realize that the first-order interacting space must be derived from single and double replacements of any type of electron -- active or inactive, the 1982 approach limited this to the inactive orbitals.
However, the new method kept the constraint adopted by the Lund group, the zeroth-order Hamiltonian should be a one-electron operator and in the case that the active space reduces to a single determinant the method should be equivalent to MP2.
Originally the method came in two flavors, the diagonal (CASPT2D) and non-diagonal (CASPT2N) (also know for a short while as CASPT2F\cite{andersson1992electric}) versions. 
However, pretty soon the former was scraped due to 
significant accuracy problems and the latter is now what we know as the CASPT2 method.
While the progress was significant, the approach also exhibited a number of problems and new issues. 
Here we can mention what is known as the intruder state problem \cite{andersson1990second,roos1995theoretical},
and the problem of computing properties of excited states -- in particular when the CASSCF reference function is of poor quality, maybe even qualitatively wrong \cite{lindh1989theoretical}.
The first issue has been addressed by various shift techniques (fixing the symptom) -- real \cite{roos1995multiconfigurational} and 
imaginary shifts \cite{forsberg1997multiconfiguration} --
but also with modifications to the partitioning of the Hamiltonian into the zeroth order Hamiltonian and 
the perturbation potential \cite{andersson1995different,ghigo2004modified} (avoiding the problem). For the latter we can mention the $\bm{g}_i$ family of modifications of the
original partitioning and the semi-empirical partitioning, the so-called IPEA shift.
Two other types of partitioning, the Dyall \cite{dyall1995choice} and much later the Fink approach \cite{fink2006two,fink_2009},
 have also been introduced to solve the intruder state problem but come at the expense of a more complicated reference Hamiltonian
 and higher computational cost.
 The Dyall partitioning is today the backbone of an alternative to the CASPT2 method -- 
 the $n$-electron valence state PT2 (NEVPT2) method \cite{angeli2001introduction,angeli2001n,Angeli_2006}. 
To address the second problem, that of computing properties of excited states, the original single-state CASPT2 method was extended to a multi-state version by 
introduction of the multi-state CASPT2 (MS-CASPT2) method \cite{finley1998multi} and variations thereof \cite{granovsky2011extended,shiozaki2011communication}.
The computational cost of the method has been significantly improved by the use of the resolution-of-the-identity approach to
avoid the explicit use of four-center two-electron integrals \cite{aquilante2008cholesky}.
During the last few years the research into CASPT2 theory has experienced a renaissance. 
A number of reports have focused on the formulation of multi-state CASPT2 \cite{li2019dynamically,battaglia2020extended,battaglia2021role}, 
alternative partitioning methods to avoid the intruder state problem (CASPT2-K) \cite{kollmar2020alternative},
or technical improvements towards reduced computational expense of the method \cite{kats2019multi,kollmar2021efficient}.

In this chapter we will present, in some detail, the mathematical expression of the method and the technical solutions to some of the 
problems experienced using the method -- suggested complementary reading can be found elsewhere \cite{lindh2020multi}. The chapter starts off with a brief introduction on the CASSCF method, to continue with a first section presenting the fundamentals of single state CASPT2 theory in the frame of the original publications.
This is followed by some specifics on the selection of the zeroth order Hamiltonian, and techniques on how to reduce the impact of intruder states.
We conclude the more formula-driven part of the chapter by a second section on the extension of the CASPT2 method to treat several states at once -- the MS-CASPT2 method.
This is followed by a third section presenting a brief summary of the performance of the method. A short exposition of further development potentials that remain to be explored can be found in section four. The chapter ends with a summary and conclusion. 

\section{Prelude: CASSCF}

Before describing the CASTP2 method in detail, it is adequate to devote a
little time to go over the basics and main features of the CASSCF method, that
provides the "zeroth-order" wave function upon which CASPT2 will try to
improve. For a more in-depth understanding of CASSCF, its development,
implementation, variants, performance and issues, the reader is advised to
refer to the publications already mentioned in the introduction
\cite{olsen2011casscf,Roos2016}, as well as the recent chapter by Li Manni et
al. \cite{LiManni2020} on multiconfigurational methods. The following is
therefore only a short summary, emphasizing the aspects relevant for the CASPT2
treatment.

CASSCF stands for complete active space self-consistent field, and it is a form
of MCSCF (multiconfigurational self-consistent field) method. The self-consistent
field method \emph{par excellence} is the Hartree--Fock method (HF), which
assumes that the wave function can be expressed as a single
Slater determinant and then proceeds to find the molecular orbitals (MO) that
minimize the expectation value of the energy of such a wave function.
Minimizing the energy means finding a stationary point of the energy function,
where the orbitals are consistent with the average field generated by all the
electrons in the system -- described by these same orbitals -- and hence the
"self-consistent field" name. In the vast majority of practical cases, the
orbitals are expressed as linear combinations of basis functions, and therefore
finding the molecular orbitals amounts to determining the matrix of MO
coefficients $\bm{C}$.

Despite its tremendous success, the HF approximation is not without problems,
and the most relevant ones are due to the restriction to a single Slater 
determinant. For instance, it is unable to provide a qualitatively correct
description of homolytic bond breaking and many open-shell systems, where 
nearly-degenerate orbitals are not fully occupied. One of the possible ways to
improve the HF description is extending the wave function \emph{ansatz} to a
linear combination of Slater determinants, including different occupied
orbitals beyond those of HF:
\begin{equation}
\label{eq:CIwf}
\Psi = \sum_i c_i\psi_i
\end{equation}
where each $\psi_i$ is a Slater determinant with the same number of electrons,
but described by a different subset of orbitals. It is well known that if this
expansion contains all possible determinants that can be built from a given set
of orbitals (spanned by a basis set), the result -- called full configuration
interaction (FCI) -- is invariant to the particular selection of orbitals. That
is, if a different set of orbitals (spanned by the same basis functions) is
generated by applying a unitary transformation, or "rotation" to the original
set ($\psi_i\to \psi'_i$), the same wave function can be expressed with an
appropriate modification of the coefficients ($c_i\to c'_i$), and thus the wave
function that minimizes the expectation value of the energy is the same,
although the numerical values of the coefficients will be different. The
situation changes when the expansion is truncated and only some of the possible
determinants are included. In this case, a change in the orbitals usually
implies that the space of possible wave functions changes and therefore the
solution that minimizes the energy depends on how the orbitals are defined or
selected. This is what happens in conventional CI methods, such as CISD
(configuration interaction with single and double excitations): a starting set
of orbitals is taken, usually from a HF calculation; a limited set of Slater
determinants is selected by choosing different occupation patterns of the
starting orbitals; and the coefficients in Eq. \eqref{eq:CIwf} are optimized.
The MO coefficients are not modified, and the obtained energy and wave function
are not invariant to them. If, on the other hand, not only the CI coefficients
$c_i$, but also the MO coefficients $\bm{C}$ are optimized, we get an MCSCF
method.

The term "complete active space" is a recipe to select the determinants or
electronic configurations that will be included in Eq. \eqref{eq:CIwf}. It
requires choosing a subset of the molecular orbitals in the system -- the
active orbitals --, a number of active electrons $n$, and generating all
possible configurations arising from distributing $n$ electrons in the active
orbitals, keeping the other orbitals either always occupied (the inactive
orbitals) or always empty (the virtual or secondary orbitals). It is thus a FCI
expansion within the subset of the active orbitals -- if all the
orbitals are selected as active, it is identical to FCI. Following the
discussion above, the term CASCI is used if only the CI coefficients are
optimized, while CASSCF if both CI and MO coefficients are optimized.

A common feature of CI and MCSCF methods is that they can be straightforwardly
applied to excited states. Instead of obtaining a single wave function, one can
define a number of orthogonal wavefunctions $\Psi_k$, each of them expressed as
a linear combination of the same set of configurations:
\begin{equation}
\label{eq:CIwf_k}
\Psi_k = \sum_i c_{ki}\psi_i
\end{equation}
where the coefficients $c_{ki}$ can be found as the eigenvectors of a
generalized eigenvalue equation:
\begin{gather}
\bm{H}\bm{c} = \bm{E}\bm{S}\bm{c} \\
H_{ij} = \left\langle \psi_i | \hat{H} | \psi_j \right\rangle \\
S_{ij} = \left\langle \psi_i | \psi_j \right\rangle
\end{gather}
with $\bm{E}$ as the energies (eigenvalues) of all the considered electronic
states. The difficulty for MCSCF methods is that the MO coefficients should be
optimized to minimize the energy, but which energy? There is now not a single
state but a number of them. One could select any specific state and minimize
the $n$th energy, but -- apart from the algorithmic problems that are likely
to appear due to the order of the energies changing in the process -- the
result would be heavily biased towards the state targeted in the optimization,
and the energies of the other states would be overestimated in comparison.
The trivial solution of repeating
the process for each of the states of interest has its own problems. For one
thing, it would mean multiplying the effort by the number of states; but more
seriously, it would break the assumption that all the states are build from the
same set of $\psi_i$, since the MO optimization for each state changes the
$\psi_i$ for that state, and as a consequence the resulting wave functions
cannot be guaranteed to be orthogonal, which in most cases is an inconvenience
that complicates analysis.

The most common way to compute excited states with MCSCF methods, and in
particular with CASSCF, is by using state averaging (SA). This means that the
energy minimized by the MO optimization is not that of any particular state,
but the average of some of them. Typically, it will be the arithmetic mean of
the lowest $n$ energies, but other schemes that give more weight to some states
are also possible. The state averaging technique avoids some of the instability
issues of the state-specific procedure, and it ensures that all the states are
expanded with the same set of configurations and therefore the wave functions
are orthogonal.

There is a critical choice when using the CASSCF method, and that is the choice
of active space. We have mentioned that it involves a number of orbitals and a
number of electrons, but we have not said anything about how these orbitals
should be chosen. The truth is that there is no definitive answer, because that
depends not only on the usual trade-off between accuracy and computational
speed, but also -- and most importantly -- on the specific system and property
one is interested in computing. A general rule of thumb is that the orbitals
that change occupation during a process, or orbitals close in energy but with
different occupations, should be active, this often includes conjugated $\pi$
orbitals in valence electronic excitations, bonding and antibonding pairs of
orbitals in bond breakings, d-shell orbitals in metallic complexes, etc.
Although the notation CASSCF($N$,$M$), where $N$ is the number of active
electrons and $M$ is the number of active orbitals, is commonly used, this is
often an incomplete description of the method, and it is crucial to identify
\emph{which} orbitals are active. The number of active electrons is often easy
to determine once the nature of the active orbitals is known.

It should be kept in mind, however, that the main purpose of a CASSCF
calculation is not to obtain very accurate wave function and properties. These,
in general, can only be attained with very large expansions, which would
require large active spaces, and the computational scaling of CASSCF is
exponential with the number of active orbitals. Instead, a good CASSCF wave
function should provide an adequate and balanced qualitative reference, that
does not suffer from the limitations of using a single electronic
configuration, and this includes the possibility of spin-adapted wave functions.
In other words, the CASSCF method is a good tool for dealing
with the static electron correlation -- arising from nearly-degenerate
configurations -- but is not the best choice for the dynamic correlation --
mostly due to short-range electron--electron repulsion. Nevertheless, there are
many cases where the required or desired active space size exceeds the
practical computational limits. To deal with these cases, one can resort to
variant methods that reduce the number of configurations in Eq. \eqref{eq:CIwf}
-- such as RASSCF (restricted active space SCF) or GASSCF (generalized active
space SCF) -- or solve the active-space FCI problem in an approximate manner --
e.g. Stochastic-CASSCF or DMRG (density matrix renormalization group). Once an
adequate reference wave function is obtained, the effect of dynamic correlation
can be incorporated with post-CASSCF methods, the most commonly used of which
is CASPT2, that will be the focus of the rest of this chapter.

\section{CASPT2 Theory}
Here we will present the basic CASPT2 theory. Initially the fundamentals as originally outlined by K. Andersson are presented. 
This follows with discussions on the most serious problem of the approach -- the intruder state problem. At this point remedies are proposed
as numerical techniques to eliminate the influence of the zero or close-to-zero energy denominators in the expression to
determine the first-order correction of the reference wave function -- shift techniques. This is then followed with a subsection on the
matter of alternative zeroth-order Hamiltonians.

\subsection{CASPT2 Fundamentals}
The development of the original CASPT2 method was cast in the light of four constraints.
First, the Lund group selected to use standard Rayleigh--Schrödinger perturbation theory (RSPT) and the intermediate
normalization approach.
Second, the unperturbed reference state should be the CASSCF wave function.
Third, the reference Hamiltonian, $\hat{H}_0$, should be an effective one-electron Hamiltonian.
Fourth and final, the CASPT2 theory should reduce to the MP2 formalism in the case the CASSCF CI expansion is reduced down to a
single Slater determinant.
It is easy to see that at least the three latter constraints will produce some complications.
The combination of the second and third constraints, for example, is problematic since the CASSCF wave function
is not known to be an eigenfunction of any natural energy operator.
Furthermore, the flexibility to tailor an effective one-electron operator is further
complicated by the fourth constraint which requires our reference Hamiltonian to reduce down to the traditional Fock operator of 
Møller--Plesset perturbation theory applied to a restricted closed-shell or open-shell Hartree--Fock Slater determinant.

We will below drive through the original developmental steps of the CASPT2 theory starting with the development of
the Hamiltonian of the reference system, the definition of the first-order interacting space and ultimately
solving the RSPT equations to establish the first-order interacting space and computing the second-order correction to
the reference energy.

\subsubsection{The $\hat{H}_0$ operator}
In the restricted closed-shell case, the Fock operator is expressed as
\begin{equation}
    \hat{H}_0 = \hat{F}= \sum_{pq} \hat{f}_{pq}
\end{equation}
where the summation indices run over the spacial atomic orbitals (not spin orbitals) and
$\hat{f}_{pq}$ is the standard molecular orbital Fock operator
\begin{equation}
    \hat{f}_{pq} = \hat{E}_{pq} \left(
    h_{pq} + \sum_{k} \left[2(pq|kk) - (pk|qk)\right]
    \right)
\end{equation}
Here the summation is over the occupied canonical SCF orbitals,
$\hat{E}_{pq}=\sum_\sigma \hat{a}^\dagger_{p_{\sigma}}\hat{a}_{q_{\sigma}}$ is the operator of the unitary group, and the matrix elements $h_{pq}$ and $(pq|rs)$ are the one-electron integrals (containing the the kinetic energy and the electron--nuclear attraction) and the electron--electron repulsion integrals in chemists notation \cite{szabo2012modern}, respectively.
The eigenvalues of the molecular orbital Fock operator, $\epsilon_p$, are known as the orbital energies, which are a measure of
the energy of an electron (the sum of the kinetic energy and the electrostatic interaction energy, including Coulombic and exchange terms) in a canonical orbital $p$ -- the corresponding eigenvector -- experience in the presence of
the Coulomb field generated by the charges of electrons of the reference state $\Psi^{(0)}$. Here the values are typically associated with
the ionization energy (IP) and the electron affinity (EA) for an occupied and a virtual orbital, respectively,
all in accordance with Koopmans' theorem \cite{koopmans1934zuordnung}.
That is, $\epsilon_p$, the orbital energy for an electron in an occupied and a virtual orbital, respectively, in the Hartree--Fock approximation, corresponds to
\begin{equation}
    \epsilon_p = \langle \Psi_N | \hat{H} | \Psi_N \rangle - \langle \Psi_{N-1}^p | \hat{H} | \Psi_{N-1}^p \rangle =  - (\text{IP})_p \qquad p \in \text{occupied orbital}
    \label{IP}
\end{equation}
and
\begin{equation}
    \epsilon_p = \langle \Psi_{N+1}^p | \hat{H} | \Psi_{N+1}^p \rangle - \langle \Psi_N | \hat{H} | \Psi_N \rangle  = -(\text{EA})_p \qquad p \in \text{virtual orbital}
    \label{EA}
\end{equation}
for normalized Slater determinants with $N$, $N-1$, and $N+1$ electrons, in which the $N-1$ and $N+1$ Slater determinants are described with the
optimized canonical orbitals of $\Psi_N$.
The superscript $p$ corresponds to the orbital from which an electron has been removed or to which it has been added for the computation of the ionization potential and the electron affinity, respectively.
We note that the molecular orbital Fock operator automatically snaps between the two cases.
If we did not have an operator like this, how would we design operators which would do the job?
We start by constructing $\Psi_{N-1}^p$ and  $\Psi_{N+1}^p$ based on the $ \Psi_N$ determinant -- expressed in the canonical SCF orbitals -- 
and the use of annihilation and creation operators, respectively, as
\begin{equation}
    \hat{a}_{p_\sigma} | \Psi_N \rangle =
    \begin{cases}
    | \Psi_{N-1}^p \rangle & p \in \text{occupied orbital} \\
    0 & p \in \text{virtual orbital}
    \end{cases}
\end{equation}
and
\begin{equation}
    \hat{a}_{p_\sigma}^\dagger | \Psi_N \rangle =
    \begin{cases}
         0 & p \in \text{occupied orbital} \\
         | \Psi_{N+1}^p \rangle & p \in \text{virtual orbital} 
    \end{cases}
\end{equation}
respectively. 
Here $\hat{a}_{p_\sigma}$ and $\hat{a}_{p_\sigma}^\dagger$ are annihilation and creation operators removing and creating an electron in 
spin orbital $p_\sigma$, respectively. We also have trivially, for wave functions represented by single Slater determinants, 
\begin{equation}
    | \Psi_N \rangle =
    \begin{cases}
          \hat{a}_{p_\sigma}^\dagger  \hat{a}_{p_\sigma} | \Psi_{N}^p \rangle & p \in \text{occupied orbital}  \\
         \hat{a}_{p_\sigma}  \hat{a}_{p_\sigma}^\dagger | \Psi_{N}^p \rangle & p \in \text{virtual orbital}
    \end{cases}
\end{equation}
We note in passing that if an annihilation operator operates on the bra representation of a wave function it
changes to a creation operator, $\langle \Psi | \hat{a}_{pq} = \langle \hat{a}_{pq}^\dagger \Psi |$, vice-versa for the creation operator.
We can now recast the operators to compute the IP and the EA in the form
\begin{equation}
    -(\text{IP})_p  = \langle \Psi_N | \hat{H} | \Psi_N \rangle - \langle \Psi_{N-1}^p | \hat{H} | \Psi_{N-1}^p \rangle 
    =  \langle \Psi_N | (\hat{a}_{p\sigma}^\dagger\hat{a}_{p\sigma} \hat{H}  - \hat{a}_{p\sigma}^\dagger \hat{H} \hat{a}_{p\sigma}) | \Psi_N \rangle
\end{equation}
which we simplify to
\begin{equation}
    -(\text{IP})_p  =  - \langle \Psi_N | (\hat{a}_{p\sigma}^\dagger [ \hat{H}, \hat{a}_{p\sigma}]| \Psi_N \rangle
\end{equation}
Correspondingly, for the electron affinity we have
\begin{equation}
    -(\text{EA})_p = \langle \Psi_{N+1}^p | \hat{H} | \Psi_{N+1}^p \rangle - \langle \Psi_N | \hat{H} | \Psi_N \rangle   
    = \langle \Psi_N | \hat{a}_{p_\sigma} \hat{H} \hat{a}_{p_\sigma}^\dagger -\hat{a}_{p_\sigma} \hat{a}_{p_\sigma}^\dagger \hat{H} | \Psi_N \rangle
\end{equation}
which we simplify to
\begin{equation}
    -(\text{EA})_p = \langle \Psi_N | \hat{a}_{p_\sigma} [\hat{H}, \hat{a}_{p_\sigma}^\dagger]| \Psi_N \rangle
\end{equation}
That is, our compound orbital Fock operator
is expressed as
\begin{equation}
    \hat{f}_{pp\sigma} =  \hat{a}_{p_\sigma} [\hat{H}, \hat{a}_{p_\sigma}^\dagger] - \hat{a}_{p\sigma}^\dagger [ \hat{H}, \hat{a}_{p\sigma}]
\end{equation}
This operator will nicely swap between computing the negative values of either the electron affinity or the ionization potential since the orbital of interest is either empty or doubly occupied, respectively,
in the case the wave function is represented by a single closed-shell Slater determinant.

For a CASSCF reference wave function a corresponding Fock-like operator will have to be slightly different in order to account for the multi-configurational and partial open-shell nature of the CASSCF wave function as well as that we are not working with some set
of canonical orbitals -- we will have non-zero off-diagonal terms. A generalized spin-averaged
orbital Fock operator is proposed of the form,
\begin{equation}
    \hat{f}_{pq} = \frac{1}{2}\sum_\sigma 
    \hat{a}_{p_\sigma}[\hat{H},\hat{a}_{q_\sigma}^\dagger]
    - \hat{a}_{p_\sigma}^\dagger[\hat{H},\hat{a}_{q_\sigma}]
    = h_{pq} + \sum_{rs} \hat{E}_{rs} \left[ (pq|rs)-\frac{1}{2}(pr|qs) \right]
\end{equation}
with the expectation value of
\begin{equation}
    f_{pq} = h_{pq} + \sum_{rs} D_{rs} \left[ (pq|rs)-\frac{1}{2}(pr|qs) \right]
    \label{eq:fock_matrix_elements}
\end{equation}
where $D_{rs}$ are elements of the 1-particle density matrix.
We note that the matrix elements of this matrix are nicely blocked in accordance with the classes of the orbitals -- inactive, active, and virtual.
Here the diagonal subblocks can be diagonalized independently, and the inactive--virtual subblocks are zero.
However, the inactive--active and active--virtual subblocks are non-zero. Thus, the diagonalization of this matrix
would correspond to eigenvectors which represents a mixing of the orbitals from the different classes.
A rotation of the orbitals to this representation would imply that the CI expansion would change to one that is not confined into any orbitals subspace but would rather include configuration state functions or Slater determinants in the full Fock space as spanned by the full-electron basis.
Initial implementations, the CASPT2D approach, ignored these off-diagonal non-zero blocks, in difference to the CASPT2N approach which included these blocks too. Empirical evidence soon demonstrated that the CASPT2D approach was inaccurate, and the CASPT2N approach is what we today know as the CASPT2 method. 
In particular, the CASPT2D method fails when occupation numbers in the active space are close to zero or two in which cases the off-diagonal coupling elements are of significance.
We proceed by expressing an $\hat{F}$ for CASPT2 theory, which is a generalization of the corresponding operator in MP2 theory,
\begin{equation}
    \hat{F} = \sum_{pq} f_{pq} \hat{E}_{pq}
    \label{Fock}
\end{equation}
This operator does not have the CASSCF wave functions as eigenfunctions.
We can, however, define a modified operator as a reference Hamiltonian which indeed has the CASSCF wave functions as eigenfunctions, namely 
\begin{equation}
    \hat{H}_0^\text{CASPT2} = \sum_{i} |\Psi_i\rangle \langle \Psi_i | \hat{F} |\Psi_i\rangle \langle \Psi_i |
    + \sum_{k,l} |\psi_k\rangle \langle \psi_k | \hat{F} |\psi_l\rangle \langle \psi_l |
\end{equation}
where $\Psi_i$ is the $i$th root of the CASSCF procedure and $\psi_k$ are electronic configurations not found in the CI space of
the CASSCF CI expansion -- for example, an electron configuration in which electrons have been moved out of the inactive orbital space or
electrons have be moved to the virtual orbital space.
An alternative expression of this is possible and preferred as we proceed towards computing the second order energy of the CASSCF state according to RSPT. 

\subsubsection{Defining the first-order interacting space}\label{define-fois}
Next, we need to compute the first-order correction to the wave function. That is, we would like to identify and manipulate the first-order interacting space. We start by dividing the configuration space into four convenient subspaces: $V_0$, $V_K$, $V_\text{SD}$, and $V_\text{TQ...}$.
The first one is the one-dimensional space of the reference CASSCF wave function. The second one is the configuration space
of all possible CASSCF CI expansions consistent with the same the number of active electrons, number of active orbitals, and
spin multiplicity as the reference CASSCF wave function but excluding this configuration.
The third subspace, $V_\text{SD}$, is the space spanned by all possible single and double electron replacements generated from $V_0$ -- for practical purposes, just as in MP2, this subspace is sometimes limited by removing virtual and/or core orbitals (the latter is the so-called frozen core approximation).
The fourth subspace is the complement to the first three subspaces.
We can alternatively express the projected referenced Hamiltonian as
\begin{equation}
\begin{split}
 \hat{H}_0^\text{CASPT2} = {}& | \Psi^{(0)}\rangle \langle \Psi^{(0)} | \hat{F} | \Psi^{(0)}\rangle \langle \Psi^{(0)} |
  +     \sum_{\substack{i,j\\i,j \in K}} |\Psi_i\rangle \langle \Psi_i | \hat{F} |\Psi_j\rangle \langle \Psi_j | + {}\\
  & \sum_{\substack{i,j\\i,j \in \text{SD}}} |\Psi_i\rangle \langle \Psi_i | \hat{F} |\Psi_j\rangle \langle \Psi_j |
  +  \sum_{\substack{i,j\\i,j \in \text{TQ...}}} |\Psi_i\rangle \langle \Psi_i | \hat{F} |\Psi_j\rangle \langle \Psi_j |
\end{split}
\label{eq:H0}
\end{equation}
which also now clearly demonstrates the block-diagonal nature of the matrix representation of the operator.
Furthermore, it is noted that with this division of the configuration space only the third subspace -- $V_\text{SD}$ -- will interact with the reference CASSCF wave function via the total Hamiltonian. The configurations in $V_K$ do not interact over the total Hamiltonian due to the variational principle -- this subspace is a null space in MP2 so here we will not find a counter example.
The $V_\text{SD}$ subspace, in general generated as $\hat{E}_{pq} \hat{E}_{rs} |\Psi_0 \rangle$, is for convenience subdivided according to the orbital classes the indices belong to. We start to note that the case where all indices are active belongs to the subspace $V_K$.
The remaining cases can be divided into 8 groups distributed over 3 families:

%\begin{table}
%   \centering
\begin{center}
    \begin{tabular}{llr}
     Internal:     & $\hat{E}_{ti} \hat{E}_{uv} |\Psi^{(0)} \rangle$     & $V_A$ \\
                  & $\hat{E}_{ti} \hat{E}_{uj} |\Psi^{(0)} \rangle$     & $V_B$ \\
     Semi-internal:& $\hat{E}_{at} \hat{E}_{uv} |\Psi^{(0)} \rangle$     & $V_C$ \\
                  & $\hat{E}_{ai} \hat{E}_{tu} |\Psi^{(0)} \rangle$ , $\hat{E}_{ti} \hat{E}_{au} |\Psi^{(0)} \rangle$     & $V_D$ \\
                  & $\hat{E}_{ti} \hat{E}_{aj} |\Psi^{(0)} \rangle$     & $V_E$ \\
     External:     & $\hat{E}_{at} \hat{E}_{bu} |\Psi^{(0)} \rangle$     & $V_F$ \\
                  & $\hat{E}_{ai} \hat{E}_{bt} |\Psi^{(0)} \rangle$     & $V_G$ \\
                  & $\hat{E}_{ai} \hat{E}_{bj} |\Psi^{(0)} \rangle$     & $V_H$
    \end{tabular}
\end{center}
%    \caption{Caption}
%    \label{tab:my_label}
%\end{table}
where $i$ and $j$ are inactive; $t$, $u$, and $v$ are active; and, $a$ and $b$ are virtual orbital indices -- this nomenclature will be maintained throughout the manuscript if not otherwise stated, $p$, $q$, $r$, and $s$ will additionally be used as general orbital indices. 
Thus, the Internal, Semi-Internal, and the External configurations have none, one, and two electrons in the virtual subspace of the orbitals, respectively.
Subspaces $V_B$, $V_E$, $V_F$, $V_G$, and $V_H$ can be further partitioned into mutually orthogonal sets by different spin couplings.
For example, in $V_G$ we can form the linear combinations
\begin{equation}
    \hat{E}_{ai} \hat{E}_{bt} |\Psi^{(0)} \rangle
    \pm \hat{E}_{bi} \hat{E}_{at} |\Psi^{(0)} \rangle
\end{equation}
with the symmetric combination corresponding to singlet coupling belonging to $V_{G^+}$
and the anti-symmetric combination corresponding to triplet coupling belonging to $V_{G^-}$.

\subsubsection{Computing the first-order interacting space and the second-order energy}
The first-order correction to the reference wave function is now expressed as a linear combination of
the configurations, $j$, in $V_\text{SD}$,
\begin{equation}
    \label{eq:first_order_wf}
    | \Psi^{(1)} \rangle  = \sum_{\substack{j\\j\in\text{SD}}}^M C_j | \Psi_j \rangle
\end{equation}
where $M \ge \dim V_{SD}$ and
the coefficients are determined by standard RSPT from the system of equations
\begin{equation}
    \sum_{\substack{j\\j\in\text{SD}}}^M C_j \langle \Psi_i | \hat{H}_0 - E_0 | \Psi_j \rangle = - \langle \Psi_i | \hat{V} | \Psi^{(0)}\rangle \qquad i=1,\ldots,M\quad i\in\text{SD}
\end{equation}
which we here recast into a matrix--vector notation as
\begin{equation}
    (\bm{H}_0 - E_0 \bm{I}) \bm{C} = - \bm{V}
    \label{eq:first}
\end{equation}
Now the second-order correction to the energy 
\begin{equation}
    E^{(2)} = \langle \Psi^{(0)} | \hat{V} | \Psi^{(1)} \rangle = \bm{V}^\dagger \bm{C}
\end{equation}
is not far away.
There is, though, a further problem, the configurations within the subclasses of $V$, as defined above, are not necessarily
linearly independent, $M > \dim V_\text{SD}$. This and near-linear dependence is removed by diagonalizing the overlap matrix, $\bm{S}_{ij}=\langle i | j \rangle$, of
the configurations in $V$ and eliminating those eigenvectors with zero or too small eigenvalues.
We first diagonalize according to
\begin{equation}
    \bm{\Lambda_S} = \bm{U}^\dagger \bm{S} \bm{U}
\end{equation}
and subsequently eliminate undesired eigenvectors in $\bm{U}$, reducing the space to $N \le \dim V_\text{SD}$.
This defines a projection from the original space to the reduced space expressed as
\begin{equation}
    \bm{\Omega} = \bm{U} \bm{\Lambda_S}^{-1/2} 
\end{equation}
Our transformed system of equations now reads
\begin{equation}
    (\bm{H}'_0 - E_0 \bm{I}) \bm{C}' = - \bm{V}'
\end{equation}
where $\bm{H}'_0= \bm{\Omega}^\dagger \bm{H}_0 \bm{\Omega}$, $\bm{C}' = \bm{\Omega}^\dagger\bm{C}$, and
$\bm{V}' = \bm{\Omega}^\dagger\bm{V}$.
This equation is solved using standard high-performance linear algebra libraries in an iterative procedure 
starting with the diagonal blocks of molecular Fock operator at the first iteration.
The second-order correction is finally computed as
\begin{equation}
    \label{eq:E2proj}
    E^{(2)} =  \bm{V}'^\dagger \bm{C}'
\end{equation}

We finally note also that while for subspaces $V_F$, $V_G$, and $V_H$ only one- and two-particle density matrices are
required for the computation of the required matrix elements to determine the first-order correction of the
wave function, the other subspaces of $V$ can require up to four-particle density matrices.
The explicit equations for the matrix elements have been published elsewhere
\cite{AnderssonThesis1992,lindh2020multi}
in some detail and will not be repeated here.

\subsection{The Intruder State Problem}
As briefly mentioned in the previous subsection, the CASPT2 approach is a methodology based
on Rayleigh--Schrödinger perturbation theory, whereby the Hamiltonian is partitioned into
a zeroth-order part, $\bm{H}_0$, and a perturbation term, $\bm{V}$, such that
\begin{equation}
    \label{eq:ISP_H_partition}
    \bm{H}(z) = \bm{H}_0 + z\bm{V}
\end{equation}
Here, we denote the perturbation strength parameter with $z$ rather than the usual $\lambda$,
to highlight the fact that in the general case, it is actually a complex number.
When $z=0$, the Hamiltonian provides a zeroth-order description of the system for which we
know the solution, while in the case $z=1$ we recover the full physical problem.
This partitioning allows us to express the total energy $E$ as an expansion around $z=0$,
which is given by
\begin{equation}
    \label{eq:ISP_RS_expansion}
    E(z) = \sum_{n=0}^{\infty} z^n E^{(n)}
    = E^{(0)} + z E^{(1)} + z^2 E^{(2)} + z^3 E^{(3)} + \ldots
\end{equation}
Even though in CASPT2 we are specifically interested in the second-order term, one might ask
the question whether this expression
%truly sums up to exact energy for $z=(1,0)$. 
converges, in general, to a finite value.
It would be particularly troublesome if the energy terms increase (in magnitude) at
each order, and the expansion ultimately diverges. Even more critical would be the case
in which $E^{(2)}$ is already so large that it does not make any physical sense;
this would completely spoil the usefulness of this approach.
Unfortunately, the latter situation does, in fact, occur in CASPT2, and it is thus very
important to understand the source of this issue.
From a phenomenological point of view, such an unphysical large contribution in the
expansion of Eq. \eqref{eq:ISP_RS_expansion} is typically due to the presence of a state
in the first-order interacting space, whose associated zeroth-order energy is degenerate
or near-degenerate with that of the reference state.
Commonly, such a state is referred to as an \emph{intruder state}, even though from
a more formal perspective, the definition of intruder state is slightly different.
In the following, we shall introduce this formal definition and explore
with a simple two-state model the convergence properties of Eq. \eqref{eq:ISP_RS_expansion}.
To do so, we will base our discussion on the excellent analysis by Helgaker, Jørgensen
and Olsen\cite{Helgaker2000}, while we point the reader to the original work by
Kato\cite{Kato1966} for a detailed account.

In general, the expansion in Eq. \eqref{eq:ISP_RS_expansion} has a finite
radius of convergence $R$, such that the series converges for $|z| < R$, and diverges
otherwise. Hence, in our case we require $R > 1$, because we are ultimately
interested in the physical problem, that is $z = 1$.
The radius of convergence is dictated by the presence of \emph{points of degeneracy}
$\zeta$, these are points in the complex plane where the energy $E(\zeta)$ takes on
the same value for two different states.
In particular, $R$ is equal to the distance from the point of expansion, $z=0$ for
Eq. \eqref{eq:ISP_RS_expansion}, to the closest point of degeneracy $\zeta$.
A state, whose energy is degenerate with that of the reference for a point $\zeta$
within the unit circle ($|\zeta| < 1$), is called an \emph{intruder state}, as
this is responsible for the divergence of the perturbation expansion.
% An intruder state for which $\Re(\zeta) > 0$ is called \emph{front-door intruder state},
% while one for which $\Re(\zeta) < 0$ is called \emph{back-door intruder state}.
It is insightful to analyze the intruder state problem using a simple two-state
model parametrized by the following Hamiltonian
\begin{equation}
    \label{eq:ISP_H_model}
    \bm{H}(z) =\bm{H}_0 + z\bm{V} =
    \begin{pmatrix}
    \alpha & 0 \\
    0 & \beta \end{pmatrix}
    + z
    \begin{pmatrix}
    0 & \delta \\
    \delta & 0 \end{pmatrix}
\end{equation}
Here, $\alpha$ is the zeroth-order energy of the reference state, and $\beta$ that
of a perturber state which interacts with the reference through the coupling $\delta$.
For this simple case we can find the analytical solution of the exact energy,
which is given by
\begin{equation}
    \label{eq:ISP_E_model}
    E_{\pm}(z) = \frac{\alpha + \beta}{2} \pm
    \frac{\sqrt{(\beta - \alpha)^2 + 4z^2\delta^2}}{2}
\end{equation}
and note that its expansion in powers of $z$ is equivalent to the expression arising
from Rayleigh--Schrödinger perturbation theory.
Importantly, Eq. \eqref{eq:ISP_E_model} can be rearranged into an expression of the form
$\sqrt{1+x}$, whose expansion in powers of $x$ is known to converge for $|x| < 1$.
We can take advantage of this fact and see that this leads to the inequality
\begin{equation}
    \frac{4\delta^2}{(\beta - \alpha)^2} < 1
\end{equation}
from which we can derive a formal condition for the convergence of the expansion
in this simple two-state model as
\begin{equation}
    \label{eq:ISP_conv_model}
    |\beta - \alpha| > 2 |\delta|
\end{equation}
In other words, Eq. \eqref{eq:ISP_RS_expansion} converges if the zeroth-order energy
difference is larger than twice the coupling between the states.
If this condition is not met, the series diverges, implying the presence of
an intruder state within the unit circle. Its position can be determined by equating
the two solutions of the energy, $E_+(z) = E_-(z)$, and solving for $z$.
This provides the location of the point of degeneracy, which comes out to be
\begin{equation}
    \label{eq:ISP_pod_model}
    \zeta_\pm = \pm \frac{\beta-\alpha}{2\delta}\mathrm{i}
\end{equation}
and is a pure imaginary number that always comes in a conjugate pair ($\zeta,\zeta^*)$.
For $|\zeta_\pm| < 1$, the perturber is an intruder.
A possible way to deal with the intruder state problem is to introduce a
\emph{level shift parameter} $\varepsilon$ that changes the energy difference between the
zeroth-order states.
This approach can be described by the following zeroth-order and perturbation matrices
\begin{equation}
    \label{eq:ISP_H_shift}
    \bm{H}_0 =
    \begin{pmatrix}
    \alpha & 0 \\
    0 & \beta + \varepsilon \end{pmatrix}
    \quad , \quad
    \bm{V} =
    \begin{pmatrix}
    0 & \delta \\
    \delta & -\varepsilon \end{pmatrix}
\end{equation}
noticing that this does not change the actual content of $\bm{H}(z)$, only its
partitioning. This can be clearly seen by computing the analytical expression for the
eigenvalues
\begin{equation}
    \label{eq:ISP_E_shift}
    E_{\pm}(z) = \frac{\alpha + \beta + (1 - z) \varepsilon}{2} \pm
    \frac{\sqrt{[(\beta - \alpha) + (1-z)\varepsilon]^2 + 4z^2\delta^2}}{2}
\end{equation}
For $z=1$, Eq. \eqref{eq:ISP_E_model} is recovered, whereas for any other
value of the perturbation strength, the level shift affects the energy. Importantly,
the latter is true also for the expansion of Eq. \eqref{eq:ISP_E_shift} in powers of $z$.
With this modified model, the point of degeneracy becomes
\begin{equation}
    \label{eq:ISP_pod_shift}
    \zeta_\pm = \frac{\beta-\alpha+\varepsilon}{4\delta^2 + \varepsilon^2}
    (\varepsilon \pm 2\delta \mathrm{i})
\end{equation}
and entails both real and imaginary components. If we now investigate the condition
for convergence $|\zeta_\pm| > 1$, we obtain the following inequality
\begin{equation}
    \label{eq:ISP_conv_shift}
    \frac{(\beta - \alpha + \varepsilon)^2}{4\delta^2 + \varepsilon^2} > 1
\end{equation}
which can be solved for $\varepsilon$. This provides a minimum value of the shift
$\varepsilon_c$, which ensures convergence of the shifted model for
\begin{equation}
    \label{eq:ISP_min_eps_shift}
    \varepsilon > \varepsilon_c = 
    \frac{4\delta^2 - (\beta - \alpha)^2}{2(\beta - \alpha)}
\end{equation}
The nice property of Eq. \eqref{eq:ISP_min_eps_shift} is that for a fixed zeroth-order
energy difference $\beta - \alpha$ and coupling $\delta$, one can always find a value
for the level shift $\varepsilon$ for which the perturbation expansion converges and
the intruder state is removed.
This, as we will see in the next section, forms the basis for practical ways to solve
the intruder state problem in CASPT2 and other multireference perturbation theory
approaches.
To conclude this analysis, we shall note that intruder states can be categorized
into two classes depending on the sign of $\Re(\zeta)$: \emph{front-door} for
$\Re(\zeta) > 0$ and \emph{back-door} for $\Re(\zeta) < 0$, respectively.
In general, front-door intruder states manifest themselves as low-lying excited states
that are nearly degenerate with the ground state, while back-door intruder states are
typically highly excited and diffuse states. The convergence pattern also differs, and
distinct \emph{archetypes} have been observed for the different types of intruder
states and partitionings\cite{Olsen2000,Olsen2019}.
For instance, front-door intruders are typically associated to an oscillating
divergent pattern, whereas back-door intruder states show an alternating
behavior of the energy corrections.
At last, while such a formal analysis on the convergence properties of perturbation
theory is important to understand its theoretical foundation, from the practical
point of view we are typically limited to just the second-order, or perhaps the
third-order correction, such that after all it is not so relevant if the expansion
ultimately diverges, as long as the low-order terms bring us closer to the exact
result.

\subsection{Shift Techniques}
From a practical standpoint, the intruder state problem appears in CASPT2 during the
solution of the first-order equations, Eq. \eqref{eq:first}, whereby the resolvent
$(\bm{H}_0 - E_0\bm{I})^{-1}$ becomes singular. This happens when $E_0$ and
\emph{at least} one eigenvalue $\epsilon_i$ of $\bm{H}_0$ are degenerate or near-degenerate.
As we have seen in the previous section, introducing a level shift $\varepsilon$
has the potential to fix this issue and
this is exactly the approach taken by \citet{roos1995multiconfigurational} with the level
shift technique.
They introduced a small real-valued parameter $\varepsilon$ in the partitioning of the
Hamiltonian according to Eq. \eqref{eq:ISP_H_shift}, and solved the CASPT2 equations
with this modified system. The resulting amplitudes of the first-order wave function,
Eq. \eqref{eq:first_order_wf}, then take on the following form
\begin{equation}
    \label{eq:ST_amplitudes}
    C_i = -\frac{
    \langle \Psi^{(0)} | \hat{V} | \Psi_i \rangle}{
    \epsilon_i - E_0 + \varepsilon}
\end{equation}
In case the original denominator becomes small or vanishes, that is
$\epsilon_i \approx E_0$, the amplitude $C_i$ does not blow up thanks to the presence
of the shift $\varepsilon$.
While these coefficients can be used directly to calculate the energy according to Eq.
\eqref{eq:E2proj}, the presence of the level shift in \emph{all} amplitudes,
even those for which $\epsilon_i - E_0 \gg 0$, is undesirable, as this unnecessarily
modifies the perturbation expansion where is not strictly needed.
To remedy this issue, a \emph{level shift correction} was devised, in which the
second-order energy contributions would be approximately equal to the unshifted ones
unless the condition $\epsilon_i - E_0 \gg \varepsilon$ is violated.
This level shift correction was derived from an analysis of the second-order energy
expression, however it turns out that it is completely equivalent to calculating the
second-order energy using Hylleraas variational expression\cite{Hylleraas1930}
\begin{equation}
    \label{eq:ST_E2var}
    E^{(2)} = \langle \Psi^{(1)} | \hat{H}_0 - E_0 | \Psi^{(1)} \rangle 
    + 2 \langle \Psi^{(0)} | \hat{V} | \Psi^{(1)} \rangle
\end{equation}
The second-order energy obtained in this way is less sensitive to the level shift
parameter than Eq. \eqref{eq:E2proj}, as shown in the original
work\cite{roos1995multiconfigurational}.

The real level shift technique has proven to be very effective to deal with intruder
states appearing in ground state calculations, however, in the calculation of
excited states or transition metal complexes the situation is different. In
these cases there is a dense manifold of electronic states close in energy, and
a larger value of the shift is typically required to avoid any intruder. Unfortunately,
this negatively affects the quality of the results. Furthermore, a closer look
to Eq. \eqref{eq:ST_amplitudes} reveals that the level shift technique does not
really \emph{remove} the singularity, but rather just moves it somewhere else.
In fact, there are many instances where the energy difference in the denominator
is negative, that can still lead to large values of the amplitudes $C_i$. This
is the case when $\epsilon_i - E_0 \approx -\varepsilon$.
To overcome these limitations, a second approach was proposed by
\citet{forsberg1997multiconfiguration}, whereby the singularity is, in principle,
completely removed. This is the \emph{imaginary level shift} technique.
The idea is to use a purely imaginary shift parameter $\mathrm{i}\varepsilon$ instead of
a real one. To avoid working with complex algebra, and considering that the final
energy should be a real-valued quantity, only the real part of the amplitudes is
considered in practice, leading to the following expression
\begin{equation}
    \label{eq:ST_amplitudes_imag}
    \Re(C_i) = - \langle \Psi^{(0)} | \hat{V} | \Psi_i \rangle
    \frac{\Delta_i}{\Delta_i^2 + \varepsilon^2}
\end{equation}
with $\Delta_i = \epsilon_i - E_0$.
Differently to Eq. \eqref{eq:ST_amplitudes}, Eq. \eqref{eq:ST_amplitudes_imag} is
guaranteed to be free of any singularity, such that no accidental intruder state
can appear due to a fortuitous match between the value of the shift and a negative
$\Delta_i$. It is interesting to note that the same shift has been developed for
the multireference MP2 approach, under the name of intruder state avoidance
technique\cite{Witek2002}.
Calculation of the second-order energy contribution with the variational expression
of Eq. \eqref{eq:ST_E2var} is once again preferred over the projected one, to
alleviate the sensitivity of the approach to the value of the shift away from the
singularities.
One crucial feature of the imaginary shift technique is its dependence on the
energy difference $\Delta_i$, in contrast to the real one that only relies on
the value $\varepsilon$ of the parameter.
It is in fact this dependence that allows the imaginary shift to completely get rid
of the singularities, albeit with a caveat.
In the discussion of the previous and current sections, we have implicitly assumed
that we have access to a diagonal zeroth-order description of the system, that is,
to the spectrum of $H_0$. While this is the case for the CASPT2D variant of the theory,
it is not affordable from a computational perspective for the full-fledged version
CASPT2N (or typically referred to as simply CASPT2). Hence, in practice, the
coefficients of Eq. \eqref{eq:ST_amplitudes_imag} are obtained using approximate
values of the energy differences, which sometime can still lead to singularities
and unphysical results.
Nevertheless, the imaginary level shift technique remains superior to the real
counterpart, and should be, in general, the preferred choice between the two.

To finish off the discussion on shift techniques, we shall reconsider for a moment
the two-state model of the previous section.
The convergence condition of Eq. \eqref{eq:ISP_conv_model} tells us that to avoid
intruder states, the zeroth-order energy difference must be larger than twice
the coupling in the perturbation potential. If this inequality is not satisfied,
there are two ways to modify it. On the one hand, one could increase the energy gap,
this is what the level shift techniques do, and on the other hand, the coupling
between the states could be reduced.
Furthermore, an additive shift to either side of Eq. \eqref{eq:ISP_conv_model} is
not the only option to modify the elements. More sophisticated functional forms
are possible, for instance exponential factors that depend on the zeroth-order
energy spectrum like the imaginary shift could have some advantages, at least
from an asymptotic perspective.
This is for instance the case of $\sigma$ and $\kappa$ regularization in MP2,
whereby singular amplitudes are suppressed or damped by an exponential factor
applied to either the energy in the denominator for the former, or the coupling
in the numerator for the latter\cite{Shee2021a}.
A similar technique could easily be applied in the context of multireference
perturbation theory as well, and in particular for CASPT2\cite{Battaglia2022}.

\subsection{Alternative Selection of the Zeroth-Order Hamiltonian}\label{Hzero}
The conventional selection of the Hamiltonian of the unperturbed system
-- the zeroth-order Hamiltonian -- has rendered us the intruder state problem.
This problem is a manifestation of that the denominator in the expression generating the
coefficients of the first order correction to the wave function, see Eq. \ref{eq:first_order_wf}, will be very small (and sometimes negative). While the previous chapter dealt with how to handle this when it occurs this chapter presents alternatives which could potentially avoid the problem altogether.
In particular, the problem is associated with the values of the matrix elements generated by $\hat{H}_0$ for the reference wave function and the configurations in the $V_\text{SD}$ space.
Hence, one could ask "Are there alternative partitionings of the Hamiltonian  which avoid this problem at the source"?  

We will below present five different definitions of the zeroth-order Hamiltonian which
try to address this. The first three are based on a one-electron operator, the two last are based on two-electron operators, this is followed by a brief summary.
However, to clarify our understanding of the problem, and to create an explicit picture in which we can understand how alternative zeroth-order Hamiltonians operate,
let us initially analyze, in some detail, the problem in terms of a CASSCF state of three electrons with a doublet spin multiplicity. 

\subsubsection{CASPT2 applied to an open-shell system}
For simplicity we
limit the orbital space to three orbitals (we denote these orbitals $i$, $t$, and $a$).
Furthermore, we design the inactive space as the $i$ orbital, now with two electrons,
and the $a$ orbital to be a virtual orbital. This leaves us with an active space of
one orbital, $t$, occupied with one electron. Thus, this CASSCF state has a single configuration state function
(CSF), which is an eigenfunction to $\hat{S^2}$ and $\hat{S}_z$. We select, as is the standard of CASSCF implementations,
that for our CSF we have $M_S=S$, that is, our normalized unperturbed wave function is described as $\Psi^{(0)}={(i)^2(t)^1}$ (two singlet-coupled electrons in $i$ and
an $\alpha$ electron in orbital $t$, hence giving rise to an overall doublet spin multiplicity of the electronic state). 
We also identify the 1-particle density matrix as
\begin{equation}
    \bm{D} = 
    \begin{pmatrix}
         2 & 0 & 0 \\
         0 & 1 & 0 \\
         0 & 0 & 0 
    \end{pmatrix}
\end{equation}
This trivially gives us the values of the orbital Fock matrix elements, using Eq. \ref{eq:fock_matrix_elements}, as
\begin{equation}
    f_{pq} = h_{pq} + 2 (pq|ii) - (pi|qi) + (pq|tt) - \frac{1}{2}(pt|qt)
    \label{eq:trivial}
\end{equation}
To proceed on that endeavor we will explore the values of the orbital Fock operator in terms of the spin-orbitals.
We start by subdividing the operator into a part that adds an electron ($+$), and one that removes an electron ($-$).
\begin{equation}
    \hat{f}_{pq} = \frac{1}{2}\sum_\sigma 
    \hat{a}_{p_\sigma}[\hat{H},\hat{a}_{q_\sigma}^\dagger]
    - \hat{a}_{p_\sigma}^\dagger[\hat{H},\hat{a}_{q_\sigma}] =  \frac{1}{2}\sum_\sigma \left( 
    \hat{f}_{pq\sigma}^{+} - \hat{f}_{pq\sigma}^{-} \right)=  \frac{1}{2}\sum_\sigma
    \hat{f}_{pq\sigma} 
\end{equation}
These correspond to evaluating the spin-averaged negatives of the electron affinity and the ionization potential, respectively.
For later convenience we will identify the matrix elements over these spin-dependent operators as the Koopmans matrices \cite{smith1975extension,day1975extension,morrell1975calculation,morrison1992extended}
 -- Dyall\cite{dyall1995choice} denotes these, after spin-averaging, as particle and hole Fock operator
 -- ,
\begin{equation}
    \begin{split}
        K_{pq\sigma}^+ & = \langle \Psi^{(0)} | \hat{f}_{pq\sigma}^{+} | \Psi^{(0)} \rangle =  \langle \Psi^{(0)} | \hat{a}_{p_\sigma}[\hat{H},\hat{a}_{q_\sigma}^\dagger] | \Psi^{(0)} \rangle\\
        K_{pq\sigma}^- &= \langle \Psi^{(0)} | \hat{f}_{pq\sigma}^{-} | \Psi^{(0)} \rangle  = \langle \Psi^{(0)} | \hat{a}_{p_\sigma}^\dagger[\hat{H},\hat{a}_{q_\sigma}] | \Psi^{(0)} \rangle 
    \end{split}
    \label{eq:koopmans}
\end{equation}
where 
the "$+$" sign indicates the energy difference between a system and the same system with one more $\sigma$ electron,
while
the "$-$" sign denotes the energy difference between a system and the same system with one less $\sigma$ electron,
respectively.
For partially occupied orbitals, a spin-averaged Fock matrix element is an average over an ionization potential and an electron affinity. 
We analyze the value of Koopmans' operators sandwiched between the reference wave function of our test case. 
\begin{equation}
   \begin{split}
    K_{ii\alpha}^- &= h_{ii} +  (ii|ii) + (ii|tt) - (it|it) = -(\text{IP})_{i_\alpha} = \epsilon_{i_\alpha}  \\
    K_{ii\beta }^- &= h_{ii} +  (ii|ii) + (ii|tt)  = -(\text{IP})_{i_\beta} = \epsilon_{i_\beta}  \\
    K_{tt\alpha}^- &= h_{tt} + 2(tt|ii) - (ti|ti) = -(\text{IP})_t\\
    K_{tt\beta }^+ &= h_{tt} + 2(tt|ii) - (ti|ti) + (tt|tt) = -(\text{EA})_t \\
    K_{aa\alpha}^+ &= h_{aa} + 2(aa|ii) + (aa|tt) - (ai|ai) - (at|at) = -(\text{EA})_{a_\alpha} = \epsilon_{s_\alpha}  \\
    K_{aa\beta }^+ &= h_{aa} + 2(aa|ii) + (aa|tt) - (ai|ai) = -(\text{EA})_{a_\beta} = \epsilon_{c_\beta}
   \end{split}
   \label{eq:28}
\end{equation}
For the $i$ orbitals we get two ionization-potential-like values with the $\alpha$ orbital extra stabilized by the spin correlation with the $\alpha$ electron in the active orbital, $t$. In the same manner we get for the $a$ orbital electron-affinity-like values with again a favorable stabilization for the $\alpha$ electron. 
Following the spin averaging we assign an ionization energy to orbital $i$, 
$f_{ii}=-\frac{1}{2}((\text{IP})_{i_\alpha}+(\text{IP})_{i_\beta})=\epsilon_i$
-- here an exchange integral $(it|it)$ is spin averaged; 
to the active orbital we assign the value $f_{tt}=-\frac{1}{2}((\text{EA})_t + (\text{IP})_t)$ --
here an electron-repulsion term seems to be averaged;
and finally we assign an electron affinity to the virtual orbital, $f_{aa}= - \frac{1}{2}((\text{EA})_{a\alpha}+(\text{EA})_{a\beta})=\epsilon_a$ - here again an
exchange integral is averaged. This turns out to be slightly inconsistent, what are we spin averaging?
To be consistent the active terms should therefore be rewritten as
\begin{equation}
   \begin{split}
    K_{tt\alpha}^- &= h_{tt} + 2(tt|ii) - (ti|ti) + (tt|tt) - (tt|tt) = -(\text{IP})_t\\
    K_{tt\beta }^+ &= h_{tt} + 2(tt|ii) - (ti|ti) + (tt|tt) = -(\text{EA})_t
   \end{split}
\end{equation}
and now the spin-averaged integral is again an exchange integral.

We note that the values derived in Eq. \eqref{eq:28} are in accordance with the use of Eqs. \eqref{eq:fock_matrix_elements} and \eqref{eq:trivial}, however,
without the detailed analysis of the origin of the values.

We continue to look at the details of the zeroth-order Hamiltonian,
\begin{equation}
    \hat{H}_0^\text{CASPT2} = | \Psi^{(0)}\rangle \langle \Psi^{(0)} | \hat{F} | \Psi^{(0)}\rangle \langle \Psi^{(0)} | + \sum_{\substack{i,j\\i,j \in K}} |\Psi_i\rangle \langle \Psi_i | \hat{F} |\Psi_j\rangle \langle \Psi_j |
\end{equation}
Let us examine the integrals in the center of each of the resolution-of-the-identity-like terms. We start with the integral for the reference
wave function. Considering that $\hat{F}$ is expressed in terms of the operator $\hat{E}_{pq}$ we note that upon projection on the
reference wave function only two diagonal terms survive,
\begin{equation}
    \langle \Psi^{(0)} | \hat{F} | \Psi^{(0)}\rangle = 2f_{ii} + f_{tt}
\end{equation}
This is exactly what we would expect, a sum of orbital energies -- just like in MP2.
However, let us be a bit more explicit and investigate the numerical values of these elements (see Eq. \eqref{eq:28}).
With the standard CASPT2 $\hat{H}_0$ operator, we surprisingly observe that the reference energy compiles to $E_0 = 2 \epsilon_i - \frac{1}{2}((\text{IP})_t + (\text{EA})_t)$
rather than $E_0 = 2 \epsilon_i - (\text{IP})_t$.
That is, through the spin averaging, CASPT2 will associate to an electron in a partially occupied
orbital an energy that contains a partial non-existing self-repulsion term, resulting in a total energy that is too high.

We now examine the matrix elements of $\hat{H}_0$ for the first-order interacting space.
For convenience the first-order interacting space is listed in Table \ref{tab:Table1}.
\begin{table}
    \centering
    \begin{tabular}{|ccc|ccc|}
        \hline
        $i$ & $t$ & $a$ & generator & subspace &$\langle \Psi^{(0)} | \hat{H}_0 | \Psi^{(0)} \rangle$, $\langle \Psi_i | \hat{H}_0 | \Psi_i \rangle$  \\
        \hline
        2 & u & 0 & \multicolumn{2}{c}{reference wave function} & $2\epsilon_i - \frac{1}{2}[(\text{EA})_t + (\text{IP})_t]$ \\
        u & 2 & 0 & $\hat{E}_{ti} \hat{E}_{tt}$ & $V_A$, internal      & $ \epsilon_i -              (\text{EA})_t - (\text{IP})_t$ \\
        2 & 0 & u & $\hat{E}_{at} \hat{E}_{tt}$ & $V_C$, semi-internal & $2\epsilon_i +                                  \epsilon_a$\\
        u & d & u & $\hat{E}_{ti} \hat{E}_{at}$ & $V_D$, semi-internal & $ \epsilon_i - \frac{1}{2}[(\text{EA})_t + (\text{IP})_t] + \epsilon_a$\\
        u & u & d & $\hat{E}_{ai} \hat{E}_{tt}$ & $V_D$, semi-internal & $ \epsilon_i - \frac{1}{2}[(\text{EA})_t + (\text{IP})_t] + \epsilon_a$\\
        u & 0 & 2 & $\hat{E}_{at} \hat{E}_{ai}$ & $V_G$, external      & $ \epsilon_i +                                  2\epsilon_a$\\
        0 & 2 & u & $\hat{E}_{ti} \hat{E}_{ai}$ & $V_G$, external      & $                          - (\text{EA})_t - (\text{IP})_t   + \epsilon_a$\\
        0 & u & 2 & $\hat{E}_{at} \hat{E}_{tt}$ & $V_H$, external      & $              -\frac{1}{2}[(\text{EA})_t + (\text{IP})_t] +2\epsilon_a$\\
        \hline
    \end{tabular}
    \caption{All possible electron configurations of three electrons in three orbitals (indexed $i$, $t$, and $a$). The
    reference state is a doublet, spin angular momentum $S=\frac{1}{2}$, with an occupation string 2u0 (legend below).
    The configurations are listed together with the generator which would generate the state relative to the reference state and
    the symbol of the subgroup in $V_\text{SD}$ to which the configuration belongs.
    Moreover, in the last column we list the values of the diagonal elements of the $\hat{H}_0$ operator.
    The occupation string uses symbols to denote: 
    2, a doubly occupied orbital;
    u, a singly occupied orbital which increases the accumulated total spin so far by a value of $\frac{1}{2}$;
    d, a singly occupied orbital which decreases the accumulated total spin so far by a value of $\frac{1}{2}$; and
    0, an empty orbital. 
    \label{tab:Table1}}
\end{table}
In general, for the matrix element between the configurations in the first-order interacting space we will have contributions from
all operators $\hat{E}_{pq}$, in particular, we will have the additional diagonal term $\hat{E}_{cc}$.
Thus, the diagonal terms, $\langle \Psi_i | \hat{H}_0 | \Psi_i \rangle$ are trivially computed, see Table \ref{tab:Table1}.
These values are an upper limit to what the values would be if $\langle \Psi_i | \hat{H}_0 | \Psi_j \rangle$ was diagonal (as in the CASPT2D
approximation). As off-diagonal elements are introduced, the eigenvalues of the matrix can reduce in value, with the possibility
that they are of the same magnitude as the reference wave function. This will be the origin of numerical instabilities and/or
singularities in Eq. \eqref{eq:first}, when we compute the coefficients of the configurations used in the first-order correction to the
reference wave function. 

At first glance we note that linear combinations of configurations in $V_C$ are potential contenders
to cause intruder state problems.
Here the diagonal term of the RHS of Eq. \ref{eq:first_order_wf} boils down to
$\epsilon_a+\frac{1}{2}[(\text{EA})_t + (\text{IP})_t]$ -- a case in which if the
electron affinity of the virtual orbital is on par with the flawed estimate of
the orbital energy of the active orbital the expression will be small or, even worse, very small.
Moreover, we note that the use of the conventional CASPT2 zeroth-order
Hamiltonian within the $V_\text{SD}$ group will cause a relative energy ordering that is potentially wrong.
The MP formalism lacks in flexibility to distinguish between removing and adding an electron to partially occupied orbitals -- we would like to have different $f_{pp}$ values depending on the case!

To sum up, the spin averaging of the Fock operator, used in the zeroth-order Hamiltonian, will underestimate the orbital
energy of spatial orbitals with partial occupation close to one electron.
This will especially be empirically evident when we compute energetics for
homolytic cleavage (binding energies), and atomization energies. 
The imbalance in the CASPT2 treatment of CASSCF wave functions dominated by a single open-shell configuration or a single closed-shell configuration
will favor the former.
To quote Andersson and Roos \cite{andersson1993multiconfigurational}
regarding the accuracy of chemical processes in which radical electron pairs are formed: 
"Heats of reactions for reactions in  which one new pair is formed are predicted to be low by between 2.5 and 6 kcal/mol". 

\subsubsection{The \textbf{g}$_i$ family of corrections}
In 1995, K. Andersson proposed three different modification to $\hat{H}_0$ such that the imbalance in the
CASPT2 treatment of closed-shell vs. open-shell electronic structures was corrected \cite{andersson1995different} -- the $\bm{g}_1$, $\bm{g}_2$ and $\bm{g}_3$  corrections.
As was noted in the previous section, some values $f_{pq}$ of the Fock matrix are a poor zeroth-order representation of the energies associated with promoting electrons to and from orbitals.
In particular, for the diagonal terms it is noted that "orbital energies" are poorly represented for, but not limited to, open-shell electronic structures due to
a spin averaging over the ionization potential and the electron affinity of that particular orbital.
Andersson proposed three different modifications to the Fock matrix following the same operational scheme, 
\begin{equation}
    \bm{f}'= \bm{f} + \bm{g}_i \qquad i=1,2,3
\end{equation}
The author analyzed the matter in a way similar to the analysis in the previous section and identified that the contributions
to the spin-averaged Fock matrix of a high-spin restricted-open-shell single-determinant reference state comes from a closed-shell part and an open-shell part.
\begin{equation}
    f_{pq} = h_{pq} + \sum_{r}^\text{closed-shell}\left[ 2(pq|rr) - (pr|qr) \right] + \sum_{r}^\text{open-shell}\left[ (pq|rr) - \frac{1}{2}(pr|qr) \right]
\end{equation}
Keeping in mind that the elements are due to the spin average over the Koopmans' matrices,
\begin{equation}
    \begin{split}
    K_{pq\alpha}^{-} &= h_{pq} + \sum_{r}^\text{closed-shell}\left[ 2(pq|rr) - (pr|qr) \right] +  \sum_{r_\alpha}^\text{open-shell}\left[ (p_\alpha q_\alpha |r_\alpha r_\alpha ) - (p_\alpha r_\alpha |q_\alpha r_\alpha ) \right] \\
    K_{pq\beta}^{+} &= h_{pq} + \sum_{r}^\text{closed-shell}\left[ 2(pq|rr) - (pr|qr) \right] +  \sum_{r_\alpha}^\text{open-shell}(p_\beta q_\beta |r_\alpha r_\alpha)
    \end{split}
\end{equation}
where the first summation is over spatial orbitals and the second one over spin-orbitals,
the terms of the 3-orbital, 1-in-1 CASSCF example from above are
\begin{equation}
    \begin{split}
    K_{bb\alpha}^{-} &= h_{bb} + 2(b b |aa) - (b a|b a) \\
    K_{bb\beta}^{+} &= h_{bb} +  2(b b|aa) - (b a|b a) + (b b |b b)
    \end{split}
\end{equation}
Andersson argues that the averaging is probably adequate for the inactive--inactive and the virtual--virtual block of the Fock matrix, however,
the active--active block would need to be modified -- averaging over the spin results in matrix elements that are poor representation of
the actual orbital energy of singly occupied orbitals. 
The author proposed that the active--active block should look more like the $K_{bb\alpha}^{-}$ part than the
average over the two spin contributions, that is, we would like to subtract an integral contribution corresponding to the
averaged exchange contribution between the electrons in general and electrons in an orbital, but only if the orbital is singly occupied.
To achieve this Andersson constructs first the so-called hole density matrix, $\bm{d}=2 \bm{I} - \bm{D}$, followed by the identification that
$\bm{D}\bm{d}$ -- a concave function -- only has non-zero elements in the active--active part. Thus,
the exchange-like integral contributions to be subtracted are represented as 
\begin{equation}
    K_{pq} = \sum_{tu} (Dd)_{tu} (pt|qu)
\end{equation}
in which the summation is limited to contributions from active orbitals and more so from singly occupied than empty or doubly occupied orbitals (note that the $K$ notation is associated with the traditional
notation for Coulomb, $J$, and exchange, $K$, integrals, respectively, and is in no way associated with
the Koopmans matrices).
The indices in $\bm{K}$ are general and need to be restricted, they should be symmetric and maybe only have non-zero elements
in the same blocks as the original $\hat{H}_0$. In this context Andersson proposed three different procedures that make such a restriction in an approximate or strict way. 
First, 
\begin{equation}
    \bm{g}_1 = - \frac{1}{4} \left[ \bm{D}\bm{K}\bm{d} + \bm{d}\bm{K}\bm{D} \right]
\end{equation}
in which $\bm{g}_1$ is symmetric and has zero inactive--inactive and virtual--virtual blocks. This correction would reduce to
exactly the sought-after correction in the case of a high-spin reference function, that is, it corrects the spin-averaged exchange term.
However, this correction also introduces non-zero elements in the otherwise zero inactive--virtual block.
Second, the author suggests
\begin{equation}
    \bm{g}_2 = - \frac{1}{2} (\bm{D}\bm{d})^{1/2} \bm{K} (\bm{D}\bm{d})^{1/2}
\end{equation}
which is perfectly zero in all blocks but the active--active block, where it is identical to the $\bm{g}_1$ correction
in the high-spin case.
Andersson notes that both suggested corrections so far have the same power dependence in $\bm{D}$ and $\bm{d}$. To break with that,
a third modification was proposed
\begin{equation}
    \bm{g}_3 = - \frac{1}{2} (\bm{D}\bm{d}) \bm{K} (\bm{D}\bm{d})
\end{equation}
which shares the properties with $\bm{g}_2$ -- only the active--active block is non-zero and for high-spin cases the correction
is identical to $\bm{g}_1$. The major difference between $\bm{g}_3$ and the other two is the power dependence which makes
the correction reduce faster as the partial occupation fades away from perfectly singly occupied orbitals.
The grand total of these approaches is that the orbital energy of the active orbitals is lowered relative to the use of the
conventional CASPT2 Fock operator. That is, the virtual--active energy difference is increased, while the inactive--active one is reduced.
The former leads to reduced contribution to the second-order energy correction from configurations associated with these orbitals, while in the case of high-energy inactive orbitals one will observe increased contributions from configurations associated with those orbitals.
Tests indicated that the corrections were essential for significant quantitative improvements -- the systematic error proportional to the number of unpaired electrons disappeared, while it could not differentiate between the three different corrections with respect to the quantitative accuracy.

\subsubsection{The IPEA shift}
Ghigo and co-workers \cite{ghigo2004modified} argued in 2004 that the $\bm{g}_i$ family of corrections were not a general solution to the intruder-state problem, rather they introduced a 
new set of intruder states since the corrections reduce the inactive--active orbital energy gap.
To correct for this they introduced a level shift in the zeroth-order Hamiltonian -- the IPEA shift.
They argue that for open-shells orbitals the zeroth order Hamiltonian should look like a
ionization potential when you excited out of the orbital, while for an excitation out of the orbital it should be
represented by an electron affinity.
To demonstrate this let us compute the matrix elements of the two Koopmans' matrices corresponding to an orbital, $p$, which is singly occupied.
We start by arbitrarily assuming the the first electron has 
an $\alpha$ spin and compute the energy corresponding to removing the electron,
\begin{equation}
    K_{pp\alpha}^{-} = h_{pp} + \sum_{rs} D_{rs} [(pp|rs) -\frac{1}{2}(pr|ps)] = -(\text{IP})_p
\end{equation}
this is followed by computing the energy of adding a second electron to the same orbital ($D_{pp}=1$),
\begin{equation}
    \begin{split}
    K_{pp\beta}^{+} & = h_{pp} + \sum_{rs} D_{rs} [(pp|rs) -\frac{1}{2}(pr|ps)] + (pp|pp) = -(\text{EA})_p \\
                & = K_{pp\alpha}^{-} + D_{pp} [2(pp|pp) - (pp|pp)] = -(\text{IP})_p + D_{pp}((\text{IP})_p - (\text{EA})_p) \\
                & = -(\text{IP})_p + D_{pp}\epsilon
    \end{split}
\end{equation}
In the case of the spin averaging the last term in the RHS, $D_{pp} [2(pp|pp) - (pp|pp)]$,
-- corresponding to a Coulomb repulsion (recast here to mimic a combined Coulomb term and an exchange term) 
with the already present electron in orbital $p$ --
is averaged over both electrons.
Hence,
the averaged Fock matrix -- especially spin-averaging the exchange integral -- for a singly occupied orbital
will attribute a non-existent repulsion energy to 
the process of removing an electron -- the exchange integral would here perfectly cancel the repulsion integral --,
a reduced repulsion energy  -- an artificial contribution from the spin-averaged exchange integral reduce parts of the repulsion --
for the process of adding the second electron.
This will lead to that for a singly occupied orbital the
energy required for removing an electron and the energy gain for the addition of a second electron is underestimated.
This will render the energy denominators in the computation of the first-order correction to the
wave function to be too small, and the associated energy contributions to be too large.
Can a formulation be achieved for singly occupied orbitals in which the removal of an electron
computes to an ionization potential, whereas the addition of a second electron amounts to
an electron affinity?
The diagonal values of the originally spin-averaged Fock matrix can alternatively be expressed as
\begin{equation}
    f_{pp} = - \frac{1}{2} \left(
    D_{pp}(\text{IP})_p + (2-D_{pp})(\text{EA})_p
    \right)
\end{equation}
The authors now propose two independent corrections to be used independently if we are
exciting into an orbital or out of an orbital, respectively.
In the former case the spin-averaged equation is
modified to 
\begin{equation}
    f_{pp} = - \frac{1}{2} \left(
    D_{pp}(\text{IP})_p + (2-D_{pp})(\text{EA})_p - D_{pp}\epsilon = (\text{EA})_p 
    \right) = (\text{EA})_p
\end{equation}
and in the latter case to
\begin{equation}
    f_{pp} = - \frac{1}{2} \left(
    D_{pp}(\text{IP})_p + (2-D_{pp})(\text{EA})_p + (2-D_{pp})\epsilon
    \right) = (\text{IP})_p
\end{equation}
This would work well if we knew $(\text{EA})_p$ and $(\text{IP})_p$. Hence,
the two working equations will be
\begin{equation}
    f_{pp} = f_{pp}^\text{spin-averaged} + \frac{1}{2} D_{pp} \epsilon
\end{equation}
and
\begin{equation}
    f_{pp} = f_{pp}^\text{spin-averaged} - \frac{1}{2}(2- D_{pp}) \epsilon
\end{equation}
respectively, with $\epsilon$ being an empirically determined parameter.
Finally, the discrimination between the two definitions of the Fock matrix is
done on the basis of the definition of the first-order interacting space,
\begin{equation}
    | \Psi_{qs}^{pr}\rangle  = | \hat{E}_{pq}\hat{E}_{rs} | \Psi^{(0)} \rangle
\end{equation}
where $p$ and $r$ are orbitals we excite into and
$q$ and $s$ are orbitals we remove electrons from. The shifts are
applied accordingly to the diagonal elements of the Fock matrix.
Thus, the IPEA shift introduces level shifts which are specific
for each subgroup of $V_\text{SD}$.
We conclude this section by emphasizing that this correction is
empirical -- the authors suggest a use of $\epsilon=\SI{0.25}{\hartree}$ as a default.
While this approach nowadays is often used, it is used
with ambiguous praise and some even argue that the correction is flawed \cite{zobel2017ipea}.

\subsubsection{Use of Koopmans matrices, CASPT2-K}
Kollmar and co-workers \cite{kollmar2020alternative} argued for yet a third way, the CASPT2-K approach, in correcting for the flaws of the spin-averaged Fock matrix and apply class-specific corrections, however, without the use of any empirical parameters.
Their approach is based on using the extended Koopmans' theorem. 
The authors establish that for multiconfigurational reference function a strict variational estimate -- under the assumption that the reference function is an eigenfunction of the Hamiltonian -- of the ionization potential and the
electron affinity can be expressed in terms of linear combinations of elements of the Koopmans matrices (see Eq. \eqref{eq:koopmans}).
Hence, they should be an excellent building block of the zeroth-order Hamiltonian.
Using the expression of the operators, already published by Dyall \cite{dyall1995choice}, they derive the explicit equations for the matrix elements for the three different orbital index classes: inactive, active, and virtual (see Eqs. 11 and 12 of ref. \citenum{kollmar2020alternative}).
This is followed by a comparison between equations of the more elaborate and strictly derived (\latin{vide infra}) Dyall Hamiltonian \cite{dyall1995choice}, $\hat{H}_\text{D}$, and the CASPT2 zeroth-order Hamiltonian \cite{andersson1990second}, $\hat{H}_0$. For example, the Dyall approach yields
\begin{equation}
    \langle \hat{E}_{t'i'} \Psi^{(0)} | \hat{H}_\text{D} - E^{(0)} | \hat{E}_{ti} \Psi^{(0)}\rangle =
    -f_{ii'}D_{t't} + \delta_{ii'}K^+_{t't}
\end{equation}
where $E^0=E_\text{CASSCF}$. The corresponding CASPT2 expression is
 \begin{equation}
    \langle \hat{E}_{t'i'} \Psi^{(0)} | \hat{H}_0 - E^{(0)} | \hat{E}_{ti} \Psi^{(0)} \rangle =
    -f_{ii'}D_{t't} + \delta_{ii'}
    \left[
    \sum_u f_{ut}D_{t'u} +
    \sum_{uv}f_{uv}
    \left(
    \Gamma_{t'tuv}+\delta_{tu}D_{t'v}-D_{t't}D_{uv}
    \right)
    \right]
\end{equation}
where $E_0 = \sum_{pq} f_{pq} D_{pq}$ and $\Gamma_{pqrs}=\langle \hat{E}_{pq}\hat{E}_{rs} \rangle - \delta_{qr}\langle \hat{E}_{ps} \rangle $
the standard 2-particle density matrix,
with similar equations for matrix elements of $| \hat{E}_{at} \Psi^{(0)}\rangle$.
Kollmar and co-workers continue and argue that the one should replace the Fock matrix elements, $f$,
in the latter so that the two expressions become identical. That is, $f \rightarrow f^+$ and
 $f \rightarrow f^-$, in the two respective cases. In the first case we have the system of linear equations expressed as follows
 \begin{equation}
     K^+_{t't} = \sum_u f^+_{ut}D_{t'u} +
    \sum_{uv}f^+_{uv}
    \left[
    \Gamma_{t'tuv}+\delta_{tu}D_{t'v}-D_{t't}D_{uv}
    \right]  
 \end{equation}
 which determine the $f^+_{uv}$ elements.
 The equation can be rewritten as
 \begin{equation}
     K^+_{tu} = \sum_{v \ge w} X^+_{u,vw} \frac{f^+_{vw}}{1+\delta_{vw}} \qquad t\ge u
 \end{equation}
 and solved with standard numerical methods. It is noted that the same Koopmans matrices occur for other excitation classes in $V_\text{SD}$.
 In what follows in the conventional CASPT2 zeroth-order Hamiltonian formalism the Fock matrix elements
 are replaced with $f^+$ and $f^-$ for excitation classes where electrons are added and removed from the active space.
 That is, for the excitation classes $\hat{E}_{ti}\hat{E}_{aj}$, $\hat{E}_{ti}\hat{E}_{uj}$, and $\hat{E}_{ti}\hat{E}_{uv}$
 $f^+_{pq}$ is used, while for $\hat{E}_{ai}\hat{E}_{bt}$, $\hat{E}_{at}\hat{E}_{bu}$, and $\hat{E}_{at}\hat{E}_{uv}$
 $f^-_{pq}$ is used.
 However, in excitation classes that leave the number of electrons in the active space intact -- $\hat{E}_{ai}\hat{E}_{bj}$, 
 $\hat{E}_{ai}\hat{E}_{tu}$, and $\hat{E}_{ti}\hat{E}_{au}$ --
 the original definition of the CASPT2 zeroth-order Hamiltonian is kept. 
 In this, the matrix elements of the projected zeroth-order Hamiltonian in $V_\text{SD}$ (see Eq. \eqref{eq:H0}), $\langle i | \hat{F} | j \rangle$, 
 are preserved for the off-diagonal elements 
 while the diagonal elements the Fock operator are modified to
 \begin{equation}
     \hat{F} \rightarrow \hat{F}^+ + \sum_{tu} (f_{tu}-f_{tu}^+)D_{tu}
 \end{equation}
 for the classes of $V_\text{SD}$ where number of electron in the active space is increased, and to
 \begin{equation}
     \hat{F} \rightarrow \hat{F}^- + \sum_{tu} (f_{tu}-f_{tu}^-)D_{tu}
 \end{equation}
 for the classes of $V_\text{SD}$ where the number of electron in the active space is decreased.
 
 Subsequent benchmarking of the CASPT2-K approach by the authors revealed accuracy improvements and reduction of the number of 
 intruder states as compared to conventional CASPT2.
 However, it was also noted, with some disappointment, that the CASPT2-IPEA approach is still better than the CASPT2-K approach, but that the latter, to its credit, is free of any empirical parameters.

\subsubsection{The zeroth-order Hamiltonians of Dyall and Fink}
The two different approaches are here discussed briefly. We start with the Dyall Hamiltonian \cite{dyall1995choice}.
Dyall discusses the shortcomings of the standard CASPT2 $\hat{H}_0$ -- the averaging
of the ionization-potential-like contribution and the electron-affinity-like contribution.
This has its origin in the use of a restricted Møller--Plesset (MP) like design of the zeroth-order
Hamiltonian. Dyall counters this approach against the Epstein--Nesbet (EN) partitioning
of the Hamiltonian \cite{epstein1926stark,nesbet1955configuration},
in which the excitation process is handled more strictly accordingly
to Koopmans' ionization potentials, and electron affinities.
In particular, Dyall notes that the MP energy denominator
in the equation for the coefficient of the first-order interacting
space is too small -- the cause of the intruder-state problem.
For the EN partitioning the denominator is increased
-- as we want -- by including corrections for the double counting
of the core repulsion and the lack of repulsion of virtual orbitals.
However, the EN partitioning also corrects for the double counting
of the core--virtual repulsion of the excited states -- this will reduce the
denominator. On the whole the empirical observation is that the latter correction
seems to dominate -- explaining the poorer convergence rate of EN theory as compared 
to MP \cite{murray1992different}.
In this respect Dyall suggests the development of a CASPT2 $\hat{H}_0$ as a combination of the MP and the EN
partitionings, employing the parts of the EN partitioning which include correction increasing the energy denominators.
In this process Dyall suggest first a hybrid $\hat{H}_0$ denoted CAS/A or even CAS/A(QD) by the introduction
of the full active-space Hamiltonian in the standard CASPT2 $\hat{H}_0$ which now is expressed as
\begin{equation}
    \hat{H}_0^\text{CAS/A} = \sum_i \hat{E}_{ii} \epsilon_i
    + \sum_{tu} h'_{tu} \hat{E}_{ii} + \frac{1}{2} \sum_{tuvw} (tu|vw) \hat{e}_{tu,vw}
    + \sum_a \hat{E}_{aa} \epsilon_a
\end{equation}
where $h'_{pq} = h_{pq} + \sum_i \left[ 2(pq|ii) - (pi|qi) \right]$ is the core Hamiltonian, and
$\hat{e}_{pq,rs} = \hat{E}_{pq}  \hat{E}_{rs} - \delta_{qr}\hat{E}_{ps}$ is a two-particle operator.
This operator is invariant to rotations of orbitals within the active space and commutes with the operator for the square magnitude of the spin, $\hat{S}^2$.
However, Dyall notes that a consistency of the treatment of the orbitals might be of importance.
There are two consistencies to consider: excitation consistency (EC) and interaction consistency (IC)
\cite{murray1992different,murray1991perturbation}.
While the first consistency is ensured by CAS/A, the second is not. Four modifications to
the CAS/A(QD) Hamiltonian are suggested -- denoted CAS/A(IC), CAS/MAX(GE), CAS/MAX(D), and CAS/MAX(AC) -- that are consistent 
with IC. We will not follow up with the details of this and advice interested readers to read the original paper.
Subsequent benchmarking by Dyall demonstrates the effectiveness of the CAS/A(QD) $\hat{H}_0$ as compared to the other 
proposed procedures. This operator is today used in the $n$-electron valence state PT2 (NEVPT2) approach \cite{angeli2001n}
which renders that approach virtually intruder-state free.
However, recent benchmarks between IPEA intruder-state corrected CASPT2 and NEVPT2 show disappointingly very similar accuracy \cite{schapiro2013assessment,loos2019reference,loos2020quest,sarkar2022assessing}, with a possible small edge for the former.

After this we continue with a short discussion of the Fink Hamiltonian, which was originally presented
in the context of SCF theory and single Slater determinants \cite{fink2006two} and later extended to a multi-configurational reference\cite{fink_2009}.

Fink defines a $\hat{H}_0$ partitioning, which he coins as the Retaining the Excitation degree Perturbation Theory (RE-PT).
Here the zeroth-order Hamiltonian is the part of the full Hamiltonian which is restricted to the components which do not
change the number of electrons in any of the orbital subspaces, i.e.,
\begin{equation}
    \hat{H}_0^\text{RE-PT} = 
    \sum_{\substack{pq\\\Delta n_\text{ex}=0}} h_{pq} \hat{E}_{pq} + \frac{1}{2} \sum_{\substack{pqrs\\\Delta n_\text{ex}=0}} (pq|rs) \hat{e}_{pq,rs}
\end{equation}
in particular the orbital indices do not run over any
combinations that will alter the number of electron in the different orbital spaces, however, the summation is not restricted to that $S_z$ in each
individual subspace is conserved.
In this respect $E^{(0)}=\langle \Psi^{(0)} | \hat{H}_0^\text{RE-PT} | \Psi^{(0)} \rangle = E_\text{CASSCF}$  
and since the reference wave function is variational, $E^{(1)}=\langle \Psi^{(0)} |\hat{H}_1^\text{RE-PT} | \Psi^{(0)} \rangle=0$.
As one proceeds with RSPT for a single determinant reference this yields an equation for the coefficients of the first-order corrections to the wave function
which is known as the equation to determine the coefficients in the linearized coupled cluster doubles (LCCD) method
\cite{vcivzek1966correlation}. In the multi-configurational version the method is called multi-reference RE (MRRE) \cite{fink_2009}.
Fink acknowledges this approach as rather computationally demanding.
To offer a remedy on this a version of the theory is proposed in which the two-electron integrals where all indices are over virtual orbitals
are Substituted with the Two-integrals  (REST-PT) over the holes in the inactive space  -- a much smaller set.
Still, however, there is a significant computational cost for any of these methods. Fink's partitioning of the Hamiltonian
has not been implemented with a multiconfigurational reference function in any
production-like code, rather performance have been assessed in association with inefficient proof-of-principle codes
\cite{sharma2015multireference,sharma2017combining,aoto2019perturbation}. 
The verdict is not out yet as to whether the RE-PT partitioning
offers any significant improvements over Dyall's zeroth-order Hamiltonian or the one in conventional CASPT2(-IPEA) theory.
Actually, it was noted that for small basis sets the RE-PT partitioning suffered from intruder-state problems \cite{aoto2019perturbation}.

\subsubsection{Summary}
The selection of an alternative $\hat{H}_0$ is an attempt to improve on conventional Møller--Plesset restricted perturbation theory, which uses spin-averaged values of the Fock operator to compute energy differences between states that differ in orbital occupation.
In this respect the ionization potential of the first electron in a doubly occupied orbital is the same
as for the second -- the first electron has an ionization energy that is too low since the repulsion energy is averaged over both electrons.
The same problem occurs when adding the first and the second electron to an otherwise empty orbital.
This flaw is at the center of the five corrections mentioned above. 
To summarize, first, Andersson's $\bm{g}_i$ family of corrections modifies the matrix elements of the Fock matrix such that the averaged
repulsion integral disappears for orbitals which are singly occupied. In this respect the ionization potential of a single electron in that
orbital is now correctly described. However, exciting a second electron into the same singly occupied orbital does now not carry the repulsion term between the electrons in the same orbital.
Regardless, Andersson uses this modified Fock matrix for all of the subgroups of $V_\text{SD}$. In particular, she notes that
new intruder states now occur due to that the inactive--active energy gap is too small. 
Second, the IPEA modification tries to introduce a similar correction as Andersson but avoiding the introduction of the second new problem.
In that respect the procedure uses an IP-like and an EA-like Fock matrix depending on if an electron is moved out of or into an orbital, respectively.
This correction term, however, contains an empirical factor, which the authors suggest should be \SI{0.25}{\hartree}.
Experience over the years is that different kinds of chemistry problems -- organic chemistry, transition metal complex stabilities, lanthanide and actinide chemistry, etc. -- operate better with larger or no IPEA shift.
Third, the Dyall $\hat{H}_0$ is a symbiosis between a MP and an EN approach. Dyall identifies for which classes the MP approach underestimates the
energy gap and replaces those with the corresponding elements offered by the EN approach, i.e. the modifications are subclass-specific.
Fourth, the Fink approach is based on an even more general partitioning of the Hamiltonian, with many similar properties to
the Dyall approach -- virtually intruder-state free, and the reference wave function is an eigenfunction of $\hat{H}_0$.
But also, the partitioning is size-consistent, and invariant to unitary rotations within each orbital subspace.
However, as noted by Fink, the new approach, which essentially in the single determinant case is consistent with the linearized coupled-cluster doubles method (LCCD), is computationally expensive.
Fifth and final, the CASPT2-K approach tries to do the same as the IPEA-shift modification, although without the use of any empirical factor.
Performance assessments of the NEVPT2 \cite{aoto2019perturbation}, CASPT2-K \cite{kollmar2020alternative}, and MRRE \cite{aoto2019perturbation} vs. CASPT2 and CASPT2-IPEA exhibit no clear difference with respect to accuracy at second order of correction. 
The major difference is in whether or not an empirical factor is employed.
To conclude, it seems that accuracy-wise a second-order multiconfigurational-reference perturbation theory approach will get no more accurate regardless of the selection of $\hat{H}_0$, the best would then be to have the simplest expression which removes the intruder state problem and at the same time is free of empirical parameters.

\section{Multi-State CASPT2 Theory}
The CASPT2 theory described above proved very successful in getting qualitatively and quantitatively
accurate results in a large number of studies, mainly in spectroscopic problems for small and medium-sized
molecules. However, there were some cases where a new limitation became evident, such as the double
crossing (instead of avoided crossing) found in the potential energy curve of LiF \cite{malrieu1995},
or the large error in the vertical excitation energy of the $1^1\text{B}_\text{1u}$ state of ethene
\cite{serrano-andres1993}. These problems arise from the fact that the CASPT2 method used so far
is applied independently to each multiconfigurational reference state, and that the second-order
treatment cannot include contributions from the orthogonal complement within the CASSCF space. i.e.
the first-order interacting space does not contain $V_\text{K}$ (see Section \ref{define-fois}).
The consequences are that even when several reference functions $\Psi_\alpha^{(0)}$ form an orthogonal
set, the corresponding first-order-perturbed functions $\Psi_\alpha^{(0)}+\Psi_\alpha^{(1)}$ are in general
not orthogonal, and that spurious mixings in the reference functions due to accidental quasi-degeneracies
at the CASSCF level cannot be resolved by the CASPT2 treatment, since the mixed CASSCF functions
are not available for the first-order interacting space.

The solution to this limitation came from the application of techniques derived from quasi-degenerate
perturbation theory (QDPT). In QDPT a set of reference functions are used to define a model space
of orthogonal functions and an effective Hamiltonian, the matrix over the reference states of the
effective Hamiltonian is then constructed and diagonalized, which gives the second-order correction
to the state energies and a set of orthogonal first-order-corrected states, potentially mixing
all the initial reference functions. The multi-state CASPT2 method (MS-CASPT2), proposed in 1998 by
Finley and co-workers \cite{finley1998multi}, differs from standard QDPT in that the $\hat{H}_0$ operator,
is not unique for all reference states, but is different for each of them. To distinguish it from
MS-CASPT2, the normal one-state CASPT2 is often called single-state, or state-specific, CASPT2 (SS-CASPT2).

To apply the MS-CASPT2 method, we start from a set of orthogonal reference states $\Psi_\alpha^{(0)}$,
produced by a single CASSCF calculation, typically a state-averaged CASSCF. The reference
states will usually be a small subset of all the possible roots of the CASSCF procedure, thus, as in SS-CASPT2,
we will still have a $V_0$ space -- spanned now by all the reference states -- and a $V_K$ space,
spanned by all the CASSCF roots that are not part of the reference states. Then, for each of the
reference states, we can build a reference Hamiltonian as in Eq. \eqref{eq:H0}:
\begin{equation}
\label{eq:H0-MS}
\begin{split}
 \hat{H}_0^\alpha = {}& \sum_\beta | \Psi_\beta^{(0)}\rangle \langle \Psi_\beta^{(0)} | \hat{F}^\alpha | \Psi_\beta^{(0)}\rangle \langle \Psi_\beta^{(0)} |
  +     \sum_{\substack{i,j\\i,j \in K}} |\Psi_i\rangle \langle \Psi_i | \hat{F}^\alpha |\Psi_j\rangle \langle \Psi_j | + {}\\
  & \sum_{\substack{i,j\\i,j \in \text{SD}}} |\Psi_i\rangle \langle \Psi_i | \hat{F}^\alpha |\Psi_j\rangle \langle \Psi_j |
  +  \sum_{\substack{i,j\\i,j \in \text{TQ...}}} |\Psi_i\rangle \langle \Psi_i | \hat{F}^\alpha |\Psi_j\rangle \langle \Psi_j |
\end{split}
\end{equation}
noting that now the $\hat{F}^\alpha$ operator is specific for each reference state, as it depends on the
state's 1-particle density matrix, see Eq. \eqref{eq:fock_matrix_elements}, and that the first
sum contains contributions from all reference states, including $\Psi_\alpha^{(0)}$. From here the
second-order energy and first-order wave function of each state is computed exactly as in the
case of SS-CASPT2. The next step is building the effective Hamiltonian matrix, i.e. the matrix with elements
$\langle \Psi_\alpha^{(0)} | \hat{H}^\text{eff[2]} | \Psi_\beta^{(0)} \rangle$, for all $\alpha$ and $\beta$ in the set of reference states. For the diagonal elements there is no ambiguity, and we
can use the corresponding SS-CASPT2 energy of each state:
\begin{equation}
H^\text{eff[2]}_{\alpha\alpha} = \langle \Psi_\alpha^{(0)} | \hat{H}^\text{eff[2]} | \Psi_\alpha^{(0)} \rangle = E_\alpha^{[2]} = E_\alpha^\text{SS-CASPT2}
\end{equation}
But for the off-diagonal elements there is an inconsistency, because different states are described
with different $\hat{H}_0$ operators. It turns out that the matrix elements can be computed as
\begin{equation}
H^\text{eff[2]}_{\alpha\beta} = \langle \Psi_\alpha^{(0)} | \hat{H}^\text{eff[2]} | \Psi_\beta^{(0)} \rangle =
\langle \Psi_\alpha^{(0)} | \hat{H} | \Psi_\beta^{(1)} \rangle \qquad \alpha \neq \beta
\end{equation}
which we can see is clearly asymmetric, $H^\text{eff[2]}_{\alpha\beta} \neq H^\text{eff[2]}_{\beta\alpha}$.
Finally, the $\bm{H}^\text{eff[2]}$ matrix is diagonalized, yielding the $E_i^\text{MS-CASPT2}$ energies as
eigenvalues. The eigenvectors describe the final MS-CASPT2 states as mixtures of the individual first-order
SS-CASPT2 states, but they can also be used to mix in the same way the reference CASSCF states, resulting
in the so-called perturbation-modified CASSCF (PM-CASSCF) states, which can be used as zeroth-order
approximations to the MS-CASPT2 states. For the diagonalization to be meaningful (real eigenvalues, orthogonal eigenvectors),
$\bm{H}^\text{eff[2]}$ should be symmetric, and the pragmatic approach of the MS-CASPT2 method is
to apply a simple symmetrization to the matrix before diagonalization:
\begin{equation}
\bm{H}^{\text{eff[2]}\prime} = \frac{1}{2} \left( \bm{H}^\text{eff[2]} + (\bm{H}^\text{eff[2]})^\dagger \right)
\end{equation}
Ideally, the off-diagonal elements are small and of similar sizes on both sides of the diagonal, so the procedure is
somewhat safe, but in cases where the values are large and different one should question the results.

The original MS-CASPT2 paper \cite{finley1998} showed that the method can solve issues like the
artificial mixing of valence and Rydberg states in ethene or spurious crossing in the potential
energy curves of LiF. In general, in cases where the CASSCF states are near-degenerate and the SS-CASPT2 states
are not, MS-CASPT2 results in a better treatment. However, when the near-degeneracy persists at the CASPT2 level,
MS-CASPT2 is known to have an erroneous behavior, particularly around conical intersections
\cite{serrano-andres2005,granovsky2011}.

This wrong behavior of MS-CASPT2 close to
conical intersections is a consequence of the use of state-specific $\hat{H}_0$ operators. Another
consequence, and a fundamental problem, is that the method is not invariant to rotations in the
reference space, i.e. unitary transformations of the initial CASSCF states give different results.
This is an especially undesirable property, again, near conical intersections (at CASSCF level), as
at an intersection any unitary transformation of the crossing states is an equally acceptable solution,
and we would not like the PT2 results to depend on which particular transformation was chosen for
the reference states. A solution for these problems was proposed in 2011 by Granovsky for the
MCQDPT method \cite{granovsky2011}, and adapted to MS-CASPT2 by Shiozaki and co-workers \cite{shiozaki2011communication}
as extended multi-state CASPT2, XMS-CASPT2.

The key differences between MS-CASPT2 and XMS-CASPT2 are that, in the latter, the Fock operator used
to define $\hat{H}_0$ is the same -- the average -- for all reference states, and that the $\hat{H}_0$
includes the full projection on the reference states, and not just the diagonal. Thus, an averaged
Fock operator is defined as in Eqs. \eqref{eq:fock_matrix_elements} and \eqref{Fock} but where
the density matrix is the average over all the $N$ reference states:
\begin{align}
D_{rs}^\text{sa} &= \frac{1}{N} \sum_\alpha D_{rs}^\alpha \\
f_{pq}^\text{sa} &= h_{pq} + \sum_{rs} D_{rs}^\text{sa} \left[ (pq|rs)-\frac{1}{2}(pr|qs) \right] \\
\hat{F}^\text{sa} &= \sum_{pq} f_{pq}^\text{sa} \hat{E}_{pq}
\end{align}
and the reference Hamiltonian is:
\begin{equation}
\label{eq:H0-XMS}
\begin{split}
 \hat{H}_0 = {}& \sum_{\alpha,\beta} | \Psi_\alpha^{(0)}\rangle \langle \Psi_\alpha^{(0)} | \hat{F}^\text{sa} | \Psi_\beta^{(0)}\rangle \langle \Psi_\beta^{(0)} |
  +     \sum_{\substack{i,j\\i,j \in K}} |\Psi_i\rangle \langle \Psi_i | \hat{F}^\text{sa} |\Psi_j\rangle \langle \Psi_j | + {}\\
  & \sum_{\substack{i,j\\i,j \in \text{SD}}} |\Psi_i\rangle \langle \Psi_i | \hat{F}^\text{sa} |\Psi_j\rangle \langle \Psi_j |
  +  \sum_{\substack{i,j\\i,j \in \text{TQ...}}} |\Psi_i\rangle \langle \Psi_i | \hat{F}^\text{sa} |\Psi_j\rangle \langle \Psi_j |
\end{split}
\end{equation}
Note the difference in the first term, compared with Eq. \eqref{eq:H0-MS}. This means
that $\hat{H}_0$ is no longer diagonal in the reference states and the reference states
are therefore not eigenfunctions of $\hat{H}_0$, which is against the requirements of
perturbation theory. This problem, however, is easily solved by diagonalizing $\hat{H}_0$
first, obtaining a set of rotated reference states $\tilde{\Psi}_\alpha^{(0)}$
\begin{equation}
| \tilde{\Psi}_\alpha^{(0)} \rangle = \sum_\beta C_{\beta\alpha} | \Psi_\beta^{(0)} \rangle
\end{equation}
With these rotated reference states the first term in Eq. \eqref{eq:H0-XMS} can be replaced by
\begin{equation}
\sum_{\alpha} | \tilde{\Psi}_\alpha^{(0)}\rangle \langle \tilde{\Psi}_\alpha^{(0)} | \hat{F}^\text{sa} | \tilde{\Psi}_\alpha^{(0)}\rangle \langle \tilde{\Psi}_\alpha^{(0)} |
\end{equation}
and we can proceed as in MS-CASPT2, but now with a common $\hat{H}_0$. It should
be noted that the $\bm{H}^\text{eff[2]}$ matrix is still non-symmetric in XMS-CASPT2,
and must be symmetrized as in MS-CASPT2.

Through the use of a unique reference Hamiltonian, XMS-CASPT2 is able to correctly describe,
at least from a qualitative point of view, regions of near-degeneracy, but it does so at the
expense of abandoning the state-specific partitioning that was so successful with CASPT2.
It is evident that SS-CASPT2 is completely insensitive (except if we change the CASSCF states)
to the number of states it is applied to, since it is applied to each state individually. For
MS-CASPT2, the partitioning of the Hamiltonian is independent of the number of states, but
the final effective Hamiltonian matrix and its diagonalization will be different for different
numbers of states. In the XMS-CASPT2, the $\hat{H}_0$ itself will depend on the number of
states included in the procedure, and we can expect that the average nature of the Fock
operator will degrade the performance of XMS-CASPT2 compared to SS-CASPT2 (and MS-CASPT2) for
states that did not suffer from the drawbacks of those methods. This observation has
recently led Battaglia and Lindh \cite{battaglia2020extended,battaglia2021role} to the proposal of a dynamically-weighted
variant, XDW-CASPT2, that recovers a state-specific partitioning, but in a way that leaves
$\hat{H}_0^\alpha$ essentially unchanged from SS-CASPT2 for states which are well separated
in energy from other states, and approaches an even average as the states become degenerate.
In the XDW-CASPT2 method there is an adjustable parameter $\zeta$ such that, on the one hand, when $\zeta = 0$
all states are averaged evenly, regardless of their energies, thus becoming equivalent to
XMS-CASPT2; on the other hand, when $\zeta \to \infty$ each state has its own completely independent $\hat{H}_0$,
as in MS-CASPT2. For intermediate values of $\zeta$ the averaging is smoothly "switched on"
for states that are close in energy. In any case, as in XMS-CASPT2, there is a pre-diagonalization
step of the reference states to obtain the rotated reference states, and this pre-diagonalization
uses an $\hat{H}_0$ built from the evenly averaged Fock operator, regardless of the value
of $\zeta$. It is only in the subsequent calculation of first-order correction to the wave
function and second-order correction to the energy that the state-specific reference Hamiltonians
are used, now based on the rotated reference states. This means that even in the case $\zeta \to \infty$
the method does not become exactly equivalent to MS-CASPT2, which involves no such rotation.
The authors call this limit case rotated multi-state CASPT2, RMS-CASPT2, and suggest it should
be used as a more robust alternative to MS-CASPT2.

\section{Performance}
Over the years, the CASPT2 method has been the subject of a number of benchmark studies
that tried to ascertain what accuracy can be expected from such calculations.
These studies have in general compared the CASPT2 results with experimental values,
with higher-level multiconfigurational methods like MRCI or Full-CI, or even with
single-reference methods like CCSDT or CC3.

In the first 1993 benchmark, Andersson and Roos \cite{andersson1993multiconfigurational} studied
35 electronic states of small molecules containing first- and second-row atoms (H to F) with a 
relatively large basis set including up to f functions. As an active space they used the
full valence set of electrons and orbitals, except for the 2s of N, O and F. The conclusion
was that bond distances are predicted with an error against experiment of \SI{0.01}{\angstrom} (somewhat larger
for very weak bonds that even CASSCF cannot properly describe), errors in the energy of
isogyric reactions (which preserve the number of electron pairs) were up to \SI{2.5}{\kcalmol},
while atomization energies were typically underestimated by \SIrange{3}{6}{\kcalmol}
\emph{per electron pair broken}. This latter fact is a manifestation of the imbalance between
singly occupied orbitals and empty or doubly occupied orbitals that led to the proposal of
the $\bm{g}_i$ family and IPEA corrections (see Section \ref{Hzero}).

A study of 11 pericyclic reactions of unsaturated hydrocarbons was published in 2003 by
Guner et al. \cite{guner2003} In this case the CASPT2 calculations were done with a smaller
basis set (6-31G*), and the active spaces where limited to 4-in-4, 6-in-6, or 8-in-8. Compared
with experimental values, the authors found that this method yields a mean absolute deviation (MAD)
of around \SI{1.7}{\kcalmol} in reaction and activation enthalpies.

Vertical excitation energies were benchmarked in 2008 and 2010 by Schreiber et al., \cite{schreiber2008,SilvaJunior2010}
by computing excitation energies to 223 singlet and triplet states of 28 medium-sized organic
molecules. In their setting, they used the TZVP and \emph{aug}-cc-pVTZ basis sets, the active space was selected depending
on each molecule and type of excitation, and the energies are based on a mixture of CASPT2-IPEA and
MS-CASPT2-IPEA calculations. The authors compare their results with previous results of the Roos
group, which were mostly based on plain CASPT2. They note that the introduction of the IPEA
correction typically increases the excitation energies, which tended to be underestimated,
by \SIrange{0.2}{0.3}{\eV}. The MAD between CC3/TZVP and CASPT2/TZVP results is \SI{0.22}{\eV}
for singlets and \SI{0.08}{\eV} for triplets. Enlarging the basis set from TZVP to \emph{aug}-cc-pVTZ
tends to decrease the excitation energies by \SIrange{0.1}{0.2}{\eV}, and the correlation
between both sets suggests that the higher the excitation energy, the more it
will be reduced by enlarging the basis set. Although some theoretical best estimates are given,
based on previously published results with MRCI, MRMP and CC with large basis, no summary statistics
are provided that compare with these theoretical results or with experiments. 

In a somewhat controversial work, Zobel et al. analyzed in 2017 \cite{zobel2017ipea} the performance
of the IPEA shift by doing both a literature survey of published results and a set of benchmark
calculations on di- and triatomic molecules. The authors conclude that, at least for organic
chromophores, the opposite effects of the IPEA shift (increase excitation energies) and
enlarging the basis set size (decrease excitation energies) lead to an error cancellation and make
the CASPT2 (no IPEA) results with double-$\zeta$ basis sets closer to experimental values than
CASPT2-IPEA results, while calculations with larger basis sets are improved by introducing the IPEA
correction. They also suggest that the optimal value for the IPEA shift will depend on the system
size and type, and that the use of a universal shift parameter (i.e. \SI{0.25}{\hartree}) may
not be justified.

More recently, Sarkar et al. \cite{sarkar2022assessing} have compared both SS-CASPT2 and SS-CASPT2-IPEA,
as well as NEVPT2, with approximations to Full CI with complete basis set, obtained from extrapolations
of coupled-cluster and selected-CI calculations \cite{Loos2018,loos2019reference,loos2020quest}.
The benchmark set included valence, Rydberg and double excitations in 35 organic molecules
containing from three to six non-hydrogen atoms. The CASPT2 calculations employed the \emph{aug}-cc-pVTZ
basis set, with an active space that depends on the molecule and excitation and typically a real
shift of \SI{0.3}{\hartree}. As expected CASPT2 without the IPEA correction is found to underestimate
excitation energies by an average of \SI{0.26}{\eV}, while CASPT2-IPEA and NEVPT2 perform very
similarly, with a MAD relative to the theoretical best estimates of \SI{0.11}{\eV} and \SI{0.13}{\eV},
respectively.

As it is clear from the above summaries, much of the interest and applications of the CASPT2 method,
in all its variants, is connected to the field of organic spectroscopy, or in general the study
of low-lying electronic excited states of small-to-medium sized molecules of light main-group atoms.
A historical perspective on the role of the CASPT2 method and B. O. Roos in this topic was given
by L. Serrano-Andrés in a special issue in memory of Prof. Roos \cite{SerranoAndres2011},
in what was probably his last contribution before he himself passed away. However, there are other
areas where the CASPT2 method has been -- and continues to be -- successfully applied, which
pose their own specific challenges.

Perhaps the most significant of these areas is that of transition metal compounds. Transition metals
are characterized by a partially filled d-shell of atomic orbitals, resulting typically in a dense
manifold of close-lying electronic states that differ in the specific occupation pattern of these
orbitals and the spin coupling of the corresponding electrons. Thus, the presence of transition metals
is often a sign of "multiconfigurationality" of the wave function, even in the ground state, and
single-reference methods tend to struggle to describe them consistently. The CASSCF reference used
by CASPT2 can be a qualitatively good approach to the wave function and so CASPT2 calculations can offer
an accuracy comparable with that of main-group molecules. An overview of how the development of
CASPT2 allowed tackling some of the problems in understanding transition metal compounds was presented
by K. Pierloot in the aforementioned special issue \cite{Pierloot2011}. One particular issue
affecting transition metal calculations, especially for top-row metals, is the so-called "double-d-shell"
effect. When considering states where the occupation of the d orbitals changes -- e.g. 3d$\to$4s or charge
transfer excitations -- a single set of d orbitals in the active space is not flexible enough
to quantitatively describe the changes in wave function and electron density, and it becomes necessary
to include a second set of d orbitals (3d$'$) to allow for some "breathing" space for the different
configurations. For the specific case of spin energetics of top-row transition metal complexes,
Pierloot et al. reported a set of benchmark calculations in 2017 \cite{Pierloot2017}, comparing
SS-CASPT2 (with different values of the IPEA shift) and NEVPT2 to MRCI and CCSD(T). They concluded that CASPT2
seems to be superior and more consistent than NEVPT2, although both methods are still erratic
in the description of the 3s3p correlation. They also noted that the results are sensitive to the
IPEA shift value, highlighting once again the desire to avoid empirical parameters.

As one moves down in the periodic table and reaches the f block (lanthanides and actinides), the
issue of multiple nearly degenerate electronic configurations is exacerbated, although the chemical
significance changes. The close-lying configurations arise now from a partially filled f-shell, which
is more compact than the d-shell and in the case of the lanthanides tends to not interact with the
ligands as much. Thus, lanthanide atoms can maintain exceptionally high spin states that are the
source of interesting magnetic and electrical properties. In the actinides, however, the f-shell
orbitals are much more exposed and typically all 5f, 6d and 7s orbitals must be included in the
active space; besides, as heavier elements, actinides manifest relativistic effects to a greater
extent than lanthanides. The importance of magnetism in lanthanides and the heavy nature of
actinides therefore demand the use of special tools and techniques that can take into account
and describe these relativistic effects, such as inclusion of scalar relativistic effects, use of
relativistically consistent basis sets, the calculation of spin--orbit coupling (e.g. through the
atomic mean-field integrals, AMFI, method \cite{Hess1996}), and the coupling of spin eigenstates
according to it (e.g. with the spin--orbit RAS state interaction, SO-RASSI, method \cite{Malmqvist2002}).
L. Gagliardi published a historical account of the early applications of CASPT2 to actinides
in the memorial issue already mentioned \cite{Gagliardi2011}.

A relatively recent field of application for computational chemistry is that of X-ray spectroscopies \cite{Norman2018}. X-rays
photons have enough energy to excite or ionize the electrons in core orbitals, giving access to a range
of spectroscopies and processes like X-ray absorption spectroscopy (XAS), resonant inelastic X-ray
scattering (RIXS) or X-ray magnetic circular dichroism (XMCD). The localized nature of the core orbitals
makes these techniques quite element-specific, but still retaining enough sensitivity to the chemical
environment to allow distinguishing between bonding patterns or oxidation states. The first challenge
one encounters when trying to apply CASPT2 to core-excited states is that these are very high in
energy, typically several hundred \si{\eV}, well above all the valence excitations or even ionized
states, this means that a straightforward application could require computing thousands of "useless"
states before the interesting ones could be reached. This problem, however, can be overcome by
a creative use of the RAS3 space of RASSCF (the core orbital to be excited from can be placed
in RAS3, with a maximum number of electrons of one), together with specific constraints that
prevent the collapse to the lower energy valence states, or with other methods that target higher-lying
solutions in the CI problem \cite{OpenMolcas}. Other challenges are the design of basis sets
with enough flexibility in the core orbitals -- typical basis sets are certainly \emph{not} optimized
for core-excited states -- or the inclusion of nuclear dynamic effects that may become crucial
in such out-of-equilibrium states.

In spite of its versatility and success, the CASPT2 is evidently not free from limitations. Some
of them have already been discussed in this chapter, along with possible solutions or mitigations --
the intruder state problem, the imbalance between open-shell and closed-shell configurations,
the treatment of near-degeneracies. But one must keep in mind that the method inherits more
fundamental limitations from its building blocks, CASSCF and second-order perturbation theory.
The use of CASSCF reference functions implies, on the one hand, that the method is not a "black box",
and that a certain amount of expertise, user input and critical analysis is necessary when building
or selecting the active space. On the other hand, it provides an additional flexibility to adapt
the calculations to the specific needs of the problem under study, the case of the double d-shell
is an example of this. Additionally, the reliance on CASSCF imposes limitations on the system
sizes (in terms of active orbitals) that can be treated in practice. The use of perturbation
theory, in turn, introduces limitations in the accuracy that can ultimately be expected from
the calculations, which should not be better than that of MP2 for well-behaved single-configuration
systems (unless the active space is enlarged to capture a significant part of the dynamic
electron correlation). The advantage of CASPT2 over MP2 is therefore not in intrinsic accuracy,
but in applicability and consistency on a wider range of chemical systems and physico-chemical challenges. From the computational
point of view, the PT2 part implies limitations not only on the size of the active space, but
also on the size of the first-order interacting space, i.e. the number of virtual orbitals and
therefore the total size of the system and basis set.

Finally, we would like to note that the CASPT2 method or some of its relatives is available
in most of the current quantum chemistry packages that implement multiconfigurational
wave function methods. The details and specific variants found in each case vary, but given
the rate of development any more specific list or description of the packages will likely be outdated
by the time this book is published. The best piece of advice we can give the interested user
is therefore to read the documentation and recent literature related to the software under consideration.

\section{Future Developments}
The activity and renewed interest in the development of CASPT2 and CASPT2-based approaches
witnessed in recent years is a clear sign of the potential of this method, even after
30 years from its inception.
In fact, thanks to ever-increasing computational resources, the use of CASPT2 and other
MRPT approaches to study systems and processes of chemical interest is more routine now
than back then. Thus, it is both crucial and timely to overcome the deficiencies of the
theory in the near future and further enlarge the range of applications that can be
targeted with it.

Two major limitations that have been undermining the accuracy and robustness of the CASPT2
approach are the unbalanced treatment of closed-shell and open-shell electronic configurations,
and the presence of intruder states.
Despite previous attempts to solve the former issue, namely the $\bm{g}_i$ family of
corrections\cite{andersson1995different}, the IPEA shift\cite{ghigo2004modified}, and the
recent CASPT2-K variant\cite{kollmar2020alternative}, we can expect the appearance of new
alternative partitionings of the Hamiltonian that try to improve on these ideas. For instance,
an approach following the IPEA and CASPT2-K philosophy of \emph{selective} modifications
of $H_0$ based on excitation classes could be envisaged, however without requiring an input
parameter or the solution of a linear system of equations.
Alternatively, completely new partitionings that still retain the one-electron nature of
the Fock operator, for instance in a manner similar to the recently developed
perturbation-adapted perturbation theory\cite{Knowles2022}, could be extended to
the multiconfigurational context.
To deal with the second issue, the intruder state problem, more sophisticated techniques
than the real and the imaginary level
shifts\cite{roos1995multiconfigurational,forsberg1997multiconfiguration} are certainly
going to appear. For instance, a viable option is the regularization of the first-order
amplitudes using an exponential factor as in regularized MP2\cite{Shee2021a}, but other 
more complicated functional forms could also be considered.
Such a renormalization scheme actually has two sides to be considered: on the one hand it can be
seen as a way to remove the singularities due to vanishing denominators, but on the
other hand it can be considered as an effective way to introduce higher-order correlation
effects that appear, for instance, in more complicated theories such as the coupled
electron pair approximation or the coupled cluster approach.

As for any other quantum chemical approach, another aspect that can be further developed
in CASPT2 is the computational efficiency. Currently, the main bottlenecks
are the calculation of high-order reduced density matrices (RDMs), the solution of the
CASPT2 equations, and, to a certain extent, the formation of the overlap and zeroth-order
Hamiltonian matrices in the first-order interacting space.
Here, several factorization techniques could be employed at different stages. Attempts
in this direction have been already used in the past with various degrees of
success\cite{Kurashige2014,Phung2016,kollmar2021efficient}.
For instance, considering that the overlap and $H_0$ matrices appearing in CASPT2 are
sparse tensors of order 4 and 6, techniques such as the Cholesky decomposition could
be used to approximate them in a controlled fashion, as done in the calculation of
electron repulsion integrals.
On the other hand, one could avoid the full calculation of expensive high-order
density matrices by means of stochastic sampling techniques\cite{Anderson2020}
or expanding the first-order wave function in a different basis than that of
internally contracted configurations. The latter option was attempted with matrix
product states in combination with Dyall's Hamiltonian\cite{Sharma2014,sharma2017combining},
however it has never been explored using the Fock operator.
While convergence in the CASPT2 equations is very sensitive to the form of the
Hamiltonian, e.g. through shifts and regularization terms, practical
ways to \emph{just} accelerate their convergence without modifying the underlying
physical description can be found in machine learning techniques,
borrowing ideas, for instance, from the accelerated coupled cluster approach
by \citet{Townsend2019}.

Another very active field in the development of multireference correlation theories is
that of approximate full CI methods. In the last 10 years many methodologies have
appeared and been used as active space solvers in the CASSCF approach, these are for
instance the density matrix renormalization group\cite{Zgid2008,Ghosh2008,Ma2017},
full CI quantum Monte Carlo\cite{LiManni2016}, several flavors of selected and incremental
configurations interaction approaches\cite{Smith2017,Zimmerman2019,Levine2020} and
even machine learning techniques\cite{Yang2020}.
All these methodologies allow to overcome the usual active space size limitation
of conventional CASSCF, and are thus extremely appealing to be used as the
zeroth-order approximation in CASPT2. For some of these methods, most notably
DMRG, this has already happened, but for many others that option remains unexplored.
The use of large active spaces
with CASPT2 has a multifaceted impact on its development and application. First, it
will increase the accuracy of the method thanks to more electrons being explicitly
correlated within a full-CI-like scheme. Second, it will shift the bottleneck of the
conventional two-step CASSCF/CASPT2 approach from the CASSCF part to the CASPT2 one,
increasing even more the need for a better computational efficiency of the approach
as discussed in the previous paragraph. Third, it will enable applications to systems
and processes that were previously out of reach, thanks to the inclusion of a larger
number of orbitals needed for their description. Fourth, all the infrastructure that
has been developed in the last decades and is available in the conventional setup,
will have to be adapted and implemented anew to take full advantage of the larger
active spaces.

At last, considering the many important experimental advances in laser technologies
and thus the increased necessity of a robust and accurate theoretical modeling of
photochemical and photophysical processes, we can expect the CASPT2 approach and
all its variants to play a crucial role in the coming years.
In particular, the quasi-degenerate extensions of CASPT2 should allow a consistent
description throughout conformational space, that is, everywhere on the potential
energy surface, but also throughout the excitation manifold, i.e. relative
energies should not be too sensitive to the input parameters (active space size,
number of states, and so on).
To explore the potential energy surfaces efficiently, analytical energy gradients
and derivative couplings are required. These have already been implemented for
a number of variants\cite{Vlaisavljevich2016,Park2017a,Nishimoto2021}, but we can
expect more to come, for instance for new formulations such as RMS-CASPT2.
Relevant for non-adiabatic dynamics, but also to give access to many magnetic
properties, is the implementation of the spin--orbit coupling operator expressed in
the basis of wave functions corrected through first-order. This is necessary to
describe inter-system crossings between surfaces of different spin multiplicities,
and forms the basis for the accurate calculation of multiplet energies, zero-field
splittings, hyperfine coupling constants, and several other properties.
Currently, this is only possible with perturbation-modified CASSCF wave functions
in combination with CASPT2 energies, and its implementation with full first-order
wave functions is just a matter of time.

To summarize, the recent revival of activity around CASPT2 and MRPT can only be
expected to continue in the near future. While major increases in its accuracy
are probably limited, most of the future development is likely going to focus
on improving its robustness, consistency, efficiency and applicability.
As both computational resources and unprecedented experimental data are bound to
increase in the coming years, many new exciting possibilities for cutting-edge
applications and developments with CASPT2 are envisaged.

\section{Summary and Conclusion}
We have here given a detailed, although not exhaustive, presentation of the CASPT2 method. After a short introduction on CASSCF, the initial chronological development of CASPT2 is described -- a strict application of RSPT to a multiconfigurational reference function. Second, the chapter addresses and explains the issue known as the intruder state problem and presents shift techniques to resolve this issue from a technical viewpoint -- how to handle the close-to-zero energy denominators of the equation to determine the
coefficients for the first order correction to the wave function. Third, alternative partitionings of the Hamiltonian are discussed and analyzed -- no conclusive evidence so far discriminates between present-day $\hat{H}_0$ operators with respect to accuracy criteria. Fourth, the multi-state versions of the CASPT2 method are presented -- revealing critical shortcomings of the original version of the theory close to conical intersections. Fifth, a short summary of benchmarks results assessing the accuracy of the approach for a wide range of chemical systems and elements in the periodic system is presented. Finally, the chapter is concluded with a discussion of the development potential of the method and what to expect in the near future as improvements and extensions of the approach.   

\begin{acknowledgement}
  S.B. acknowledges the Swiss National Science Foundation (SNSF) for the funding
  received through the Postdoc Mobility fellowship.
  R.L. acknowledges the Swedish Research Council (VR, Grant No. 2020-03182), and would also like to thank Kerstin Andersson for numerous interesting discussions which shed significant light on the early developments of the CASPT2 method.
\end{acknowledgement}

\bibliography{main}

\end{document}